\documentstyle[12pt]{article}
\newtheorem{lemma}{Lemma}
\newtheorem{theorem}{Theorem}
\newtheorem{proposition}{Proposition}
\newcommand{\nab}{\overline{\nabla}}
\newcommand{\g}{\overline{g}}
\newcommand{\R}{\overline{R}}
\newcommand{\n}{{\bf{n}}}
\newcommand{\r}{{\ell}}
\newcommand{\Gam}{\tilde{\Gamma}}
\newcommand{\otheta}{\underline{\theta}}
\newcommand{\1}{\underline{1}}
\begin{document}
\title{Geometry of General Hypersurfaces in Spacetime: Junction Conditions.}

\author{Marc Mars and Jos\'e M. M. Senovilla\\ Departament de F\'{\i}sica
Fonamental, Universitat de Barcelona\\ Diagonal 647, 08028 Barcelona, Spain
\\and\\
Laboratori de F\'{\i}sica Matem\`atica, Societat Catalana de F\'{\i}sica, IEC.}
\date{}

\maketitle

\begin{abstract}
We study imbedded {\it general} hypersurfaces in spacetime i.e.  hypersurfaces
whose timelike, spacelike or null character can change from point to point.{\it
In\-he\-ri\-ted} geometrical structures on these hypersurfaces are defined by
two distinct methods: the first one, in which a {\it rigging vector} (a vector
not tangent to the hypersurface anywhere) induces the standard {\it rigged
connection}; and the other one, more adapted to physical aspects, where each
observer in spacetime induces a completely new type of connection that we call
the {\it rigged metric connection} which is volume preserving.  The
generalisation of the Gauss and Codazzi equations are also given. With the
above machinery, we attack the problem of matching two spacetimes across a
general hypersurface. It is seen that the {\it preliminary junction conditions}
allowing for the correct definition of Einstein's equations in the
distributional sense reduce to the requirement that the first fundamental form
of the hypersurface be continuous, because then, there exists a maximal $C^1$
atlas in which the metric is continuous. The Bianchi identities are then proven
to hold in the distributional sense.  Next, we find the {\it proper junction
conditions} which forbid the appearance of singular parts in the curvature.
These are shown equivalent to the existence of coordinate systems where the
metric is $C^1$. Finally, we derive the physical implications of the junction
conditions: only six independent discontinuities of the Riemann tensor are
allowed. These are six matter discontinuities at non-null points of the
hypersurface. For null points, the existence of two arbitrary discontinuities
of the Weyl tensor (together with four in the matter tensor) are also allowed.
The classical results for timelike, spacelike or null hypersurfaces are
trivially recovered.
\end{abstract}
\newpage
\section{Introduction.}
The purpose of this paper is twofold. First, we wish to find the possible
geometrical structures that {\it general} hypersurfaces inherit from the
spacetime manifold. And second, we study the application of these results to
the matching of spacetimes across one of those general hypersurfaces.

Here, {\it general} hypersurface means an imbedded three-dimensional manifold
without specifying anything about its timelike, spacelike or null character,
which will also be permitted to {\it change from point to point}. In principle,
it might seem that general hypersurfaces in the above sense are not very
physical, given that physical particles travel along timelike lines which will
never become null, least of all spacelike. But this is a very simplified naive
view, and in fact general hypersurfaces are commonplace in General Relativity.
A few examples are: G\"{o}del's universe, where there are no imbedded
hypersurfaces without boundary which are spacelike everywhere \cite{HE}; the
apparent horizon of Vaidya's radiating metric \cite{I2} (whenever the mass
function becomes a constant after a while); or even something as simple as the
stationary limit surface of Kerr's vacuum solution \cite{HE}, which is
everywhere timelike except at points on the axis where it is null and tangent
to the horizon. The reader can also find his/her own preferred examples.

The trouble with general hypersurfaces is that they inherit metrics from the
spacetime (the first fundamental form) which can be, at some points,
degenerate, and also the signature is not constant. Therefore, they do not have
an intrinsic inherited Riemannian structure, and the usual metric connection
cannot be defined in them. The degeneracy of the first fundamental form appears
also in null hypersurfaces, but for general hypersurfaces we have the change of
signature as an added problem. Imagine, for example, a hypersurface which is
spacelike in some open set, then changes to a null hypersurface. In the
spacelike part, we might be tempted to define the canonical Riemannian
structure defined by the non-degenerate first fundamental form, but then this
structure (the connection) blows up as we approach the region where the
hypersurface is null. For null hypersurfaces we do not have this problem, as
the metric is degenerate everywhere.  Thus, we see that, even in the case of a
general hypersurface with just one single point of different signature, we
cannot define the affine structure which would be good enough for the
hypersurface without that point.

The study of degenerate metrics has been addressed since long ago, but usually
keeping the signature (or the degeneracy) constant. The question is simply how
to define a good connection in general hypersurfaces which is somehow induced
by the affine structure of the whole manifold. The first references we are
aware of are those of Bortolotti \cite{Bo1},\cite{Bo2}, where he studied
absolutely specialized degenerate metrics, or in more simple terms, metrics
with non-zero eigenvectors with zero eigenvalue and vanishing Lie derivative
along these vectors.  A little later, Hlavat\'y
\cite{H} defined canonical and unique induced connections for general hypersurfaces
in which the second fundamental form is non-degenerate. This is also explained
in the famous Schouten book \cite{S}, which is an unavoidable reference for all
these matters, and where the rigged connection we define in Sect.3 is
thoroughly studied. This type of rigged connection has also been used later
several times with different purposes (see, for example,
\cite{V},\cite{Le},\cite{D1},\cite{D2},\cite{R1},\cite{R2},\cite{B}). The paper by
Bonnor \cite{B} is specially interesting in the sense that there appears the
possible importance of using observers to define the rigged connections for the
first time.  We shall elaborate on this idea in Sect.4, where we provide a
completely new way of inducing a connection onto general hypersurfaces. We call
this new connection the rigged metric connection, and we claim that it is
physically reasonable. By means of this connection, we shall also be able to
define an induced volume element in the hypersurface which is preserved by the
rigged metric connection, something which was not possible with the old rigged
connections in general.

The second part of this paper deals with the junction conditions which must be
imposed when the matching hypersurface is a general one, where we apply the
results developed in the first part. As far as we know, this problem has not
been considered previously. The junction conditions for time- or space-like
hypersurfaces are known since the work of Darmois
\cite{Da}, Lichnerowicz \cite{L} and O'Brien-Synge \cite{OS} and
the relations between them were established by Bonnor \& Vickers
\cite{BV} (see also Bel \& Hamoui \cite{BH}). More recently,
the junction conditions for null hypersurfaces have also been studied by
several authors \cite{T},
\cite{CD}, and
a unified treatment of the three types of hypersurfaces has been produced as
well in \cite{BI},\cite{Ba}. However, the problem of generalizing the junction
conditions to arbitrary hypersurfaces, which we carry out in Sect.5, is of most
interest in Gravitation. Consider, for example, the question of phase
transitions in the early universe, which can take place at spacelike regions,
the information being transmitted from there on by causal (null) signals. The
resulting whole hypersurface of discontinuous change is thus formed by a
spacelike region and the boundary of its causal future, which is null. Yet
another example. Imagine that we wish to match the Kerr vacuum solution to some
interior spacetime and that the matching hypersurface turns out to be, {\it
precisely}, the stationary limit surface. In all these and other similar cases,
we need to know the proper junction conditions for general hypersurfaces.

Finally, in Sect.6 we complete our study by deducing the physical implications
of the junction conditions, that is, the allowable discontinuities in the
matter contents and in the pure gravitational field once a proper matching has
been done. The continuity of the normal components of the energy-momentum
tensor, usually called Israel's conditions, were known for the case of non-null
hypersurfaces since the work by Israel \cite{I}. The generalization of these
conditions to the case of null hypersurfaces appears in the recent paper by
BarrabŠs and Israel \cite{BI}. Here, we present in the last section the
generalization of these conditions to arbitrary matching hypersurfaces. We also
give the continuity properties of the general Riemann tensor and of the Weyl
tensor, the latter representing the pure gravitational field. To our knowledge,
these conditions were previously unknown, despite of its evident interest for
problems involving shock gravitational waves, or even for traditional cases of
timelike or spacelike matching hypersurfaces. In this last case, we prove that
the possible Weyl discontinuities are forced by those in the matter contents,
the latter being the truly independent allowed discontinuities, while at null
points of a general hypersurface arbitrary discontinuities not related to the
matter contents are possible in two of the ten independent components of the
Weyl tensor.
\section{General Definitions and Basic Results.}
We will consider throughout this paper an oriented four-dimensional Riemannian
manifold $\left(V_4,g\right)$, with signature $+2$, and a hypersurface defined
in it. More strictly, let us also have an orientable three-dimensional manifold
$\Sigma$, and a $C^3$ map $\Phi$ from this manifold to $V_4$
\begin{eqnarray*}
\Phi : \Sigma & \rightarrow & V_4 \\
         \xi  & \rightarrow & \Phi\left(\xi\right) \equiv x\left(\xi\right) ,
\end{eqnarray*}
which is an imbedding \cite{HE} of $\Sigma$ into $V_4$. Locally, the
hypersurface $\Phi\left(\Sigma\right) \in V_4$ can be defined by a function $F$
from $V_4$ to the real numbers, ${\rm I\!R}$, through the equation $F(x)=0$.

As usual, we can construct the pull-back of covariant tensors in $V_4$ and the
push-forward of contravariant tensors in $\Sigma$. For any point $p \in
\Sigma$, the imbedding $\Phi$ allows us to write the differential map (or
push-forward) from the tangent plane of $\Sigma$ at $p$,
$T_p\left(\Sigma\right)$, to the tangent plane of $V_4$ at
$\Phi\left(p\right)$, $T_{\Phi\left(p\right)}\left(V_4\right)$,
\begin{eqnarray*} 
\left. d\Phi \right\vert_p: T_p\left(\Sigma\right) & \rightarrow &
T_{\Phi\left(p\right)}\left(V_4\right) \\ \vec{V}   & \rightarrow &
\left. d\Phi \right\vert_p \left(\vec{V}\right),
\end{eqnarray*}
which is of rank 3 at any $p \in \Sigma$, and its generalization to
contravariant tensors of any order in $\Sigma$.  Similarly, the pull-back maps
the dual tangent plane of $V_4$ at $\Phi\left(p\right)$,
$T^{\star}_{\Phi\left(p\right)}\left(V_4\right)$, onto the dual tangent plane
of $\Sigma$ at $p$, $T^{\star}_p\left(\Sigma\right)$,
\begin{eqnarray*}
\left. \Phi^{\star} \right\vert_p: T^{\star}_{\Phi\left(p\right)}\left(V_4\right) 
& \rightarrow & T^{\star}_p\left(\Sigma\right) \\
\mbox{\boldmath$\omega$}& \rightarrow & 
\left. \Phi^{\star} \right\vert_p\left(\mbox{\boldmath$\omega$}\right),
\end{eqnarray*}
and it is again of maximum rank. This map can also be extended to covariant
tensors of any order in the manifold $V_4$.

Because of the rank of the push-forward, we have that $\left. d\Phi
\right\vert_p\left(T_p\left(\Sigma\right)\right)$ is a three-dimensional linear
subspace of $T_{\Phi\left(p\right)}\left(V_4\right)$ and it is the tangent
plane to the hypersurface in $\Phi\left(p\right)$. We will denote this tangent
plane at $\Phi\left(p\right)$ as $T_p\Sigma$. Let us take a coordinate system on
$\Sigma$, $\left \{ \xi^a \right \} $ where $a$ runs from 1 to 3, in a
neighbourhood of a point $p\in \Sigma$, and a coordinate system on $V_4$,
$\left \{ x^{\alpha} \right \}$ where $\alpha$ goes from 0 to $3$, in a
neighbourhood of $\Phi\left(p\right) \in V_4$. The three vectors
$\left.\frac{\partial}{\partial\xi^a}\right\vert_p$ constitute a basis for
$T_p\left(\Sigma\right)$ and the push-forward maps them into three linearly
independent vectors at $\Phi\left(p\right)$ which are a basis for the tangent
plane to the hypersurface. Therefore we can write
\begin{eqnarray*}
\left. d\Phi \right\vert_p\left(\left.\frac{\partial}{\partial\xi^a}\right\vert_p
\right)=\left.\frac{\partial\Phi^{\mu}}{\partial\xi^{a}}\frac{\partial}
{\partial x^{\mu}}\right\vert_{\Phi\left(p\right)} \equiv
\left. e^{\mu}_a\frac{\partial}{\partial x^{\mu}}\right\vert_{\Phi\left(p\right)}
\equiv\left.{\vec{e}}_a\right\vert_{\Phi\left(p\right)}.
\end{eqnarray*}
As vector fields, ${\vec{e}}_a$ are defined only on the hypersurface
$\Phi\left(\Sigma\right)$. Given that the map $\Phi$ is an homeomorphism
between $\Sigma$ and $\Phi\left(\Sigma\right)$, we will from now on identify
the points $p$ and $\Phi\left(p\right)$ and the sets $\Sigma$ and
$\Phi\left(\Sigma\right)$ in order to simplify the notation. We can consider
the orthogonal complement of the tangent plane $T_p\Sigma$ in the dual space
$T^{\star}_p\left(V_4\right)$, which is obviously a one-dimensional linear
subspace. This one-dimensional subspace is generated by a non-zero one-form at
$p$ that we will denote by $\left.\n\right\vert_p$, which is uniquely defined
up to a non-zero multiplicative factor $\sigma\left(p\right)$, and is called
the normal form of the hypersurface or simply the normal to the hypersurface.
We define the normal vector to the hypersurface as the vector obtained by
raising the index of $\n$ with the metric of $V_4$.  As a consequence of its
definition, $\n$ is only defined on $\Sigma$.  In the coordinate basis
$\n=n_{\mu}dx^{\mu}$ and we have
$\left.\n\left({\vec{e}}_a\right)\right\vert_p=0$ or, in components,
$\left.n_{\mu}e_a^{\mu}\right\vert_p=0$.

The fact that $V_4$ is a Riemannian manifold with metric tensor $g$ allows us
to define uniquely a symmetric two-covariant tensor in $\Sigma$ by using the
pull-back.  This symmetric tensor $\Phi^{\star}\left(g\right)$ will be called
$\g$ and is the {\it first fundamental form} of the hypersurface.  In the basis
$\left\{d\xi^{a}\right\}$ the components of $\g$ are $\g_{ab}=g_{\alpha\beta}
e_a^{\alpha}e_b^{\beta}$ and of course it is defined only on the hypersurface.
There is also another two-covariant tensor $K$ on $\Sigma$ defined as $
K=\Phi^{\star}\left(\nabla\n\right)$, where $\n$ is any extension of the
one-form field $\n$ outside the hypersurface. The definition of $K$ is
independent of this extension, as is evident from the expression of its
components in a coordinate system
\[ K_{ab}=e_a^{\mu}e_b^{\nu}\nabla_{\mu}n_{\nu} \enspace .\]
This tensor on $\Sigma$ is obviously symmetric and is called the {\it second
fundamental form} of the hypersurface.

Using the canonical volume form $\mbox{\boldmath$\eta$}$ of $V_4$ one can find
an explicit expression of the normal $\n$ in terms of the basis vectors of the
tangent plane to the hypersurface. In components we have
\begin{eqnarray}
n_{\mu}=A^{-1}\eta_{\mu\alpha\beta\gamma}e^{\alpha}_1e^{\beta}_2e^{\gamma}_3
\label{ene},
\end{eqnarray}
where $A$ is an arbitrary scalar function on $\Sigma$, different from zero
everywhere, which reflects the freedom that exists in choosing the normal to
the hypersurface.  A simple calculation shows that the norm of the normal form
is
\begin{equation}
n_{\mu}n^{\mu}=-A^{-2}\det\left({}^3{\g}\right ) \label{nor},
\end{equation}
where ${}^3{\g}$ is the determinant of the first fundamental form on the
hypersurface. So we can write the following very well known result.
\begin{lemma}
At a point $p \in \Sigma$ the first fundamental form is degenerate if and only
if the normal vector is null at $p$.
\end{lemma}
We define the volume element 3-form on $\Sigma$, $\eta_{abc}$, by
\begin{eqnarray}
\eta_{123}\,n_{\alpha}\equiv\eta_{\alpha\beta\gamma\delta}e^{\beta}_1
e^{\gamma}_2e^{\delta}_3 \label{vol}
\end{eqnarray}
or, equivalently, $\eta_{abc}=A\delta_{abc}$ where $\delta_{abc}$ is the
standard alternating symbol of Levi-Civita. Last expression shows that this
volume element depends on the normalization factor of $n_{\alpha}$ and we see
that fixing by any means the volume element on the hypersurface is equivalent
to choosing the normalization factor of the normal form. We define also the
contravariant volume element on $\Sigma$ as $\eta^{abc}\equiv
A^{-1}\delta^{abc}$ in order to satisfy the usual property in Riemannian
manifolds: $\eta_{abc}\eta^{def}=\delta_{abc}^{def}$ where $\delta_{abc}^{def}$
is the Kronecker tensor.

Let $p$ be any point in the hypersurface, the tangent vectors to the
hypersurface at $p$ can be uniquely characterized as the vectors $\vec{V}
\in T_p\Sigma$ such that $\left. \n\right\vert_p\left(\vec{V}\right)=0$.
Therefore $\left. \vec{n}\cdot\vec{n}\right\vert_p=0 $ is equivalent to $\left.
\vec{n}\right\vert_p \in T_p\Sigma$, where $\vec{n}$ is the normal vector to
the hypersurface and the dot means scalar product with the metric in $V_4$.
For a point $p$ in the hypersurface we can consider the set of vectors in
$T_p\left(V_4\right)$ that are orthogonal to the tangent plane $T_p\Sigma$.  So
we define
\[ \perp T_p\Sigma \equiv \left\{ \vec{V} \in T_p\left(V_4\right) ;\ \
 g\left(\vec{V},\vec{Y}\right)=0 \ \ \forall \ \vec{Y} \in T_p\Sigma \right\}
=<\!\vec{n}_p\!> \enspace .\] This set is obviously a one-dimensional linear
subspace of $T_p\left(V_4\right)$ and it is generated by $\left.
\vec{n}\right\vert_p$. We have already seen that these two linear subspaces,
$T_p\Sigma$ and $\perp T_p
\Sigma$, will have non-zero vectors in common if and only if the
normal vector at $p$ is null. So we can write the following
\begin{lemma}
$<\!\vec{n}_p\!> \cap T_p\Sigma = \left\{\vec{0}\right\}
\Leftrightarrow \left. \vec{n}\cdot\vec{n}\right\vert_p\not=0
\Leftrightarrow T_p\left(V_4\right)
=<\!\left.\vec{n}\right\vert_p\!> \oplus T_p \Sigma$.
\end{lemma}  
Here the second equivalence follows immediately from the first one because
$<\!\left.\vec{n}\right\vert_p\!>$ is one-dimensional and $T_p\Sigma$ is
three-dimensional.

Let us now briefly recall the usual case of hypersurfaces whose normal vector
is not null at any point, i.e. $\vec{n}\cdot\vec{n}\not=0$ everywhere on
$\Sigma$ and by continuity the sign of $\vec{n}\cdot\vec{n}$ must be constant
on the whole hypersurface. By Lemma 1 we know that the first fundamental form
on the hypersurface is not degenerate and then $\Sigma$ is a Riemannian
manifold that, in consequence, possesses a unique connection associated with
the metric.  We will find explicitly this connection in a way that can be
easily generalised to the case of general hypersurfaces.  We have at any point
$p \in \Sigma$ the decomposition of the tangent plane
$T_p\left(V_4\right)=<\!\left.\vec{n}\right\vert_p\!> \oplus T_p \Sigma$ and
then any vector $\vec{V}\in T_p\left(V_4\right)$ can be decomposed uniquely
into its parallel and its orthogonal part $
\vec{V}=\vec{V}_{\perp}+\vec{V}_{\|} $ where the parallel component
$\vec{V}_{\|} \in T_p\Sigma$ and the orthogonal component $\vec{V}_{\perp} \in
<\!\left.\vec{n}\right\vert_p\!>$.  As a consequence of standard results in the
theory of dual spaces we can decompose the dual tangent plane as
$T^{\star}_p\left(V_4\right) = <\!\left.\n\right \vert_p\!> \oplus
A^{\perp}_p$, where $<\!\left. \n \right \vert_p\!>$ is the linear space
orthogonal to $T_p\Sigma$ (in the sense of dual spaces, not of the metric)
which is obviously generated by the normal one-form $\n$, and $A^{\perp}_p$ is
the 3-dimensional linear subspace orthogonal to
$<\!\left.\vec{n}\right\vert_p\!>$ defined as $A^{\perp}_p \equiv \left\{
\mbox{\boldmath$\omega$} \in T^{\star}_p\left(V_4\right) ;\,
\mbox{\boldmath$\omega$}\left(\left.\vec{n}\right\vert_p\right)=0 \right\}$.
With this decomposition of the tangent plane we can define a map $T$ from the
whole tangent plane $T_p\left(V_4\right)$ onto the tangent plane to the
hypersurface $T_p \Sigma$ by assigning to any vector in the tangent plane its
component parallel to the hypersurface
\begin{eqnarray*}
T: T_p\left(V_4\right) & \rightarrow & T_p\Sigma \\
\vec{A} & \rightarrow & \vec{A}_{\|}
\end{eqnarray*}
This map is linear and has rank 3 at any point $p \in \Sigma$. Considering the
definitions of the pull-back and the normal form $\n$ we have that
Ker$\left(\Phi^{\star}\right)=<\!\left.\n\right\vert_p\!>$.  The decomposition
of the dual tangent plane at the point $p$ and the fact that the the rank of
the pull-back is 3 allows us to establish that $\Phi^{\star}$ is an isomorphism
between $A^{\perp}_p$ and $T^{\star}_p\Sigma$. Therefore there exists an
inverse map, that we will call $\Lambda$, from $T_p\Sigma$ onto $A^{\perp}_p$
which assigns to any one-form on the hypersurface, $\mbox{\boldmath$\Omega$}
\in T^{\star}_p\Sigma$, the unique one-form on the manifold,
$\Lambda\left(\mbox{\boldmath$\Omega$}\right) \in T^{\star}_p
\left ( V_4 \right )$ with the properties
$\Phi^{\star}\left(\Lambda\left(\mbox{\boldmath$\Omega$}\right)\right)=
\mbox{\boldmath$\Omega$} $
and $ \left [\left.\Lambda\left (\mbox{\boldmath$\Omega$}\right )\right
]\left(\vec{n}\right)\right\vert_p=0$.  These two maps, $T$ and $\Lambda$, can
be respectively generalised to act on contravariant and covariant tensors of
any order.

From now on and for the sake of simplicity in the notation, we will use the
same symbol to denote a vector (or vector field) tangent to the hypersurface
considered as a vector in the manifold $V_4$ or as a vector in the
three-dimensional manifold $\Sigma$.  Let us then consider two vector fields
$\vec{X}$ and $\vec{Y}$ defined on the hypersurface and tangent to it
everywhere, that is to say: $\forall p \in \Sigma$, $\left.\vec{X}\right\vert_p
\, ,\left .\vec{Y}\right\vert_p\in T_p \Sigma$.  The vector field
$\nabla_{\vec{X}}\vec{Y}$ is well defined on the hypersurface in the sense that
there is no need of extending $\vec{X}$ or $\vec{Y}$ out of the hypersurface in
order to calculate it.  However, $\nabla_{\vec{X}}\vec{Y}$ can have, in
general, a non-zero orthogonal component. Discarding this orthogonal component
we obtain the operation $\nab_{\vec{X}}\vec{Y}$ defined as
$\nab_{\vec{X}}\vec{Y}\equiv T\left(\nabla_{\vec{X}}\vec{Y}\right)\equiv
\left(\nabla_{\vec{X}}\vec{Y}\right)_{\|}$ which
is a covariant derivative without torsion on the hypersurface. The standard
proof of this result can be found in \cite{HE},\cite{S}, and it is important to
note that this proof makes use nowhere of the fact that the vector field
$\vec{n}$ is the normal to the hypersurface. However, this is the key point in
proving the second property of this connection, namely: $\nab$ is the unique
metric connection associated with the metric $\g$ of the hypersurface.

As a final remark regarding this metric connection, let us mention that a very
simple calculation shows \cite{HE},\cite{S} that its Riemann tensor,
$R^{a}_{bcd}$, verifies the two following well-known relations called {\it the
Gauss equation}\,:
\begin{eqnarray*}
e_{d\mu}R^{\mu}_{\alpha\beta\gamma}e^{\alpha}_ae^{\beta}_be^{\gamma}_c=
R^f_{abc}{\g}_{fd}-\frac{1}{\vec{n}\cdot\vec{n}}K_{bd}K_{ca}+
\frac{1}{\vec{n}\cdot\vec{n}}K_{cd}K_{ba}   
\end{eqnarray*}
which is obviously independent on the normalization of $\n$, and {\it the
Codazzi equation}\,:
\begin{eqnarray*}
n_{\mu}R^{\mu}_{\alpha\beta\gamma}e_a^{\alpha}e_b^{\beta}e_c^{\gamma}=
\nab_cK_{ba}-\nab_bK_{ca}-\frac{1}{2\left (\vec{n}\cdot\vec{n}\right )}
K_{ba}\nab_c\left(\vec{n}\cdot\vec{n}\right)+\frac{1}{2\left
(\vec{n}\cdot\vec{n}
\, \right)}K_{ca}\nab_b\left(\vec{n}\cdot\vec{n}\right).
\end{eqnarray*}
With the usual normalization $\vec{n}\cdot\vec{n}=\pm1$, the last two terms of
this equation vanish and the Codazzi equation takes the standard, more
simplified, form.

Let us now return to the case of general hypersurfaces and try to generalise
the previous construction. The main fact that has allowed us to define a
covariant derivative on the hypersurface was the decomposition of the tangent
plane $T_p\left(V_4\right)$ at any point $p\in \Sigma$. In the general case,
however, it is not true that the normal vector $\vec{n}$ does not belong to the
tangent plane $T_p\Sigma$ for every $p$, so we cannot follow exactly the same
steps as before. To avoid this difficulty, let us define a {\it rigged}
hypersurface \cite{S} as a hypersurface $\Sigma$ where we have taken a vector
field which does not belong to the tangent plane $T_p \Sigma$ anywhere.  This
vector field, $\vec\r\,$, called the {\it rigging}, is defined only on $\Sigma$
and, obviously, it can be chosen in many different ways.  The question now is
to find out the structure that the riggings induce on $\Sigma$ and then try to
fix one (or some) of them with specially desirable properties.

Given a rigging $\vec\r\,$, we can decompose the tangent plane at every point
$p \in \Sigma$ as $T_p\left(V_4\right)=<\!\left.\vec\r\,\right\vert_p\!>\oplus
\, T_p \Sigma$ and therefore, analogously as before, the dual tangent plane at
$p$ is decomposed as
$T^{\star}_p\left(V_4\right)=<\!\left.\n\right\vert_p\!>\oplus \,
A_p^{\vec\r}$, where $A_p^{\vec\r}$ is the dual orthogonal to
$<\!\left.\vec\r\,\right\vert_p\!>$. It is evident that $A^{\vec\r}_p$ is a
three-dimensional linear subspace of $T_p^{\star}\left(V_4\right)$ which
depends on the particular choice of the rigging $\vec\r$. As before, we can
define the linear maps $T$ and $\Lambda$ and its generalizations to tensors of
any order. The decomposition written above does not change if we multiply the
rigging by a factor depending on the point of the hypersurface and, due to the
fact that $\n\left({\vec\r}\,\right)$ must be different from zero, we can
always choose this factor such that $\n\left({\vec\r}\,\right)=1$ everywhere on
$\Sigma$.  Then, the vector fields $\left\{\vec\r,{\vec{e}}_a\right\}$
constitute a basis of the tangent planes to $V_4$ at any point on $\Sigma$ and
the dual basis is given by $\left\{\n,\mbox{\boldmath $\omega$}^a \right\}$
satisfying
\begin{eqnarray*}
\r^{\alpha}\omega^a_{\alpha}=0, \hspace{0.3cm}
\omega^a_{\alpha}e_b^{\alpha}=\delta_b^a \enspace , \hspace{0.3cm} 
n_{\alpha}e_a^{\alpha}=0, \hspace{0.3cm} n_{\alpha}\r^{\alpha}=1 \, .
\end{eqnarray*}   
The pull-back and push-forward and the maps $T$ and $\Lambda$ can be made
explicit when considered in that basis as follows.  First of all, let $\Xi$ be
an arbitrary covariant tensor field in $V_4$ whose components in the coordinate
basis $\{dx^{\alpha}\}$ are $\Xi_{\alpha_1\cdots
\alpha_q}$. The pull-back of this tensor is a covariant tensor on
the hypersurface with components in the basis $\{d\xi^a\}$
\begin{eqnarray*}
{\left [\Phi^{\star}\left(\Xi\right)\right]}_{a_1\cdots
a_q}=\Xi_{\gamma_1\cdots\gamma_q} e_{a_1}^{\gamma_1}e_{a_2}^{\gamma_2}\cdots
e_{a_q}^{\gamma_q} \enspace .
\end{eqnarray*}
Similarly, for an arbitrary contravariant tensor on the hypersurface,
$\Upsilon$, with components $\Upsilon^{a_1\cdots a_r}$ in the basis
$\left\{\frac{\partial}{\partial\xi^a}\right\}$, the push-forward gives a
contravariant tensor in $V_4$ with components in the coordinate basis
$\left\{\frac{\partial}{\partial x^{\alpha}}\right\}$
\begin{eqnarray*}
{\left [d\Phi\left(\Upsilon\right)\right]}^{\gamma_1\cdots\gamma_r}=
\Upsilon^{a_1\cdots a_r}e_{a_1}^{\gamma_1} e_{a_2}^{\gamma_2}\cdots
e_{a_r}^{\gamma_r} \enspace .
\end{eqnarray*}  
The map $T$ assigns to any contravariant tensor on $V_4$, say $\Theta$, with
components $\Theta^{\gamma_1\cdots\gamma_r}$ in the basis
$\left\{\frac{\partial}{\partial x^{\alpha}}\right\}$ a contravariant tensor on
the hypersurface which, in the basis
$\left\{\frac{\partial}{\partial\xi^{a}}\right\}$, has the following components
\begin{eqnarray*}
{\left[T\left(\Theta\right)\right]}^{a_1\cdots a_r}=\Theta^{\gamma_1
\cdots\gamma_r}\omega^{a_1}_{\gamma_1}
\omega^{a_2}_{\gamma_2}\cdots\omega^{a_r}_{\gamma_r} \enspace . 
\end{eqnarray*} 
Finally, for an arbitrary covariant tensor in the hypersurface, $\Delta$, whose
components in the basis $\{d\xi^{a}\}$ are $\Delta_{a_1\cdots a_q}$, the map
$\Lambda$ produces a covariant tensor in the manifold $V_4$ with components in
the basis $\{dx^{\alpha}\}$
\begin{eqnarray*}
{\left[\Lambda\left(\Delta\right)\right]}_{\gamma_1\cdots\gamma_q}=\Delta_{a_1
\cdots a_q}\omega^{a_1}_{\gamma_1}
\omega^{a_2}_{\gamma_2}\cdots\omega^{a_q}_{\gamma_q} \enspace .
\end{eqnarray*}
From these four expressions we observe the intrinsic definition of the
pull-back and push-forward, and the dependence of $T$ and $\Lambda$ on the
rigging $\vec{\r}$.  Some particular cases of these relations concerning the
rigging and normal vectors and that we will use later in this paper are
\begin{eqnarray}
& {\left[T \left( \vec{n}\right) \right ]}^a \equiv n^a=n^{\alpha}
\omega^a_{\alpha} \enspace ,
& {\left[{\Phi}^{\star}\left(\mbox{\boldmath $\r$}\,\right)\right ]}_a \equiv
{\r}_a=\r_{\alpha}e_a^{\alpha} \enspace , \nonumber \\ &
{\left[{\Phi}^{\star}\left({\n}\right)\right ]}_a \equiv n_a=
n_{\alpha}e_a^{\alpha}=0 \, , & {\left[T \left( \vec{\r}\,\right) \right ]}^a
\equiv \r^a=\r^{\alpha} \omega_{\alpha}^a=0 \, ,
\label{var} \\
& n^a\r_a=1-\left( \vec{n}\cdot\vec{n} \right ) \left (\vec{\r}\cdot\vec{\r} \,
\right ) \, , \hspace{1cm} & 
{\g}_{ab}n^b=-\left( \vec{n}\cdot\vec{n} \right )\r_a \enspace . \nonumber
\end{eqnarray}
 
Given an arbitrary tensor field in $V_4$, defined at least on the hypersurface,
we can define another tensor field in $V_4$, defined {\it only} on $\Sigma$, by
transporting it first into the hypersurface and then back towards the manifold.
If the tensor, say $W$, has components $W_{\gamma_1\cdots
\gamma_q}^{\beta_1\cdots\beta_r}$, the image tensor, denoted $\tilde{W}$, will
have components
\begin{eqnarray*}
\tilde{W}_{\gamma_1\cdots\gamma_q}^{\beta_1\cdots\beta_r}=W_{\rho_1\cdots
\rho_q}^{\delta_1\cdots\delta_r}e_{a_1}^{\rho_1}\omega^{a_1}_{\gamma_1}
e_{a_2}^{\rho_2}\omega^{a_2}_{\gamma_2}\cdots
e_{a_q}^{\rho_q}\omega^{a_q}_{\gamma_q}
e_{b_1}^{\beta_1}\omega^{b_1}_{\delta_1}
e_{b_2}^{\beta_2}\omega^{b_2}_{\delta_2}\cdots
e_{b_r}^{\beta_r}\omega^{b_r}_{\delta_r}
\end{eqnarray*}
as can be easily checked. The object $P_{\beta}^{\gamma}\equiv
e_{a}^{\gamma}\omega^{a}_{\beta}$ appears here in a natural way.  Using the
decomposition of the unit tensor
$\delta^{\gamma}_{\beta}=\r^{\gamma}n_{\beta}+e_1^{\gamma}{\omega}_{\beta}
^{1}+e_2^{\gamma}{\omega}_{\beta}^{2}+e_3^{\gamma}{\omega}_{\beta} ^{3}$ we
find the explicit expression
\begin{equation}
P_{\beta}^{\gamma}\equiv  e^{\gamma}_a\omega^{a}_{\beta}
=\delta_{\beta}^{\gamma}-n_{\beta}\r^{\gamma} \enspace . \label{proj}
\end{equation}
The following properties show that $P_{\beta}^{\gamma}$ is the projection
tensor to the hypersurface (with respect to the rigging)
\begin{eqnarray*}
P_{\beta}^{\gamma}e^{\beta}_a=e^{\gamma}_a \, ,\hspace{5mm}
P_{\beta}^{\gamma}\r^{\beta}=0, \hspace{5mm} P_{\beta}^{\gamma}n_{\gamma}=0,
\hspace{5mm} P_{\beta}^{\gamma}\omega^{a}_{\gamma}=\omega^{a}_{\beta} \,
,\hspace{5mm} P_{\beta}^{\gamma}P_{\delta}^{\beta}=P_{\delta}^{\gamma} \,
,\hspace{5mm} P_{\gamma}^{\gamma}=3 \, .
\end{eqnarray*}
Thus, $\tilde{W}$ is the complete projection to the hypersurface of $W$ (with
respect to the rigging) in the sense that
\begin{eqnarray*}
\r^{\gamma_i}{\tilde{W}}^{\beta_1\cdots\beta_r}_{\gamma_1\cdots\gamma_q}=0
\hspace{5mm} 
\forall \, i=1 \ \dots \ q \, , \hspace{1cm}
n_{\beta_j}{\tilde{W}}^{\beta_1\cdots\beta_r}_{\gamma_1\cdots\gamma_q}=0
\hspace{5mm}
\forall \, j=1 \ \dots \ r \enspace .
\end{eqnarray*}
\section{First Connection in a General Hypersurface:
 The Rigged Connection.}

In this section we generalize, to the case of general
hypersurfaces, the results seen in the previous section for non-null
hypersurfaces. Let us consider then three vector fields $\vec{X},\vec{Y}$ and
$\vec{Z}$ on $\Sigma$, which are tangent everywhere to the hypersurface, and
let us construct the operator $\nab_{\vec{X}}\vec{Y}\equiv
T\left(\nabla_{\vec{X}}\vec{Y}\right)\equiv
{\left(\nabla_{\vec{X}}\vec{Y}\right)}_{\|}$ where now the parallel part is
taken with respect to the decomposition of the tangent plane defined by the
rigging vector $\vec\r$. We have again the following result \cite{S}:
\begin{theorem}
For each rigging, the operation $\nab_{\vec{X}}\vec{Y}$ is a torsion-free
covariant derivative on $\Sigma$.
\end{theorem}
The proof follows exactly the same steps than that of non-null hypersurfaces
mentioned above. Nevertheless, in this case we cannot prove that this
connection is metric with respect to the first fundamental form, as is obvious
because we do not have in general a canonical non-degenerate metric on the
hypersurface.

We call this connection the {\it rigged connection}. Let us find now an
explicit expression for its Christoffel symbols in the coordinate basis
${\vec{e}}_a$.  First of all we note that if we decompose the covariant
derivative $\nabla_{\vec{X}}\vec{Y}$ into its parallel part and $\vec\r$-part
and use the definition of the second fundamental form we find
\begin{eqnarray}
\nabla_{\vec{X}}\vec{Y}={\left(\nabla_{\vec{X}}\vec{Y}\right)}_{\|}-K\left(
\vec{X},\vec{Y}\right)
\vec\r \enspace . \label{dec}
\end{eqnarray}
But the Christoffel symbols are defined by $\nab_{{\vec{e}}_a}{\vec{e}}_b=
\Gamma_{ba}^{c}{\vec{e}}_c$ and then from the definition of the covariant 
derivative $\nab$ and the above expression we immediately get
$\Gamma^{c}_{ba}{\vec{e}}_c=\nabla_{{\vec{e}}_a}{\vec{e}}_b+K\left({\vec{e}}_a,
{\vec{e}}_b\right)\vec\r$, and contracting now with
$\mbox{\boldmath$\omega$}^{c}$ we obtain
\begin{eqnarray}
\Gamma^{c}_{ba}=\omega^{c}_{\mu}e^{\nu}_a\nabla_{\nu}e^{\mu}_b \, , \hspace{2cm}
\Gamma^c_{ab}=\Gamma^c_{ba} \enspace .\label{conex}
\end{eqnarray}
We will now relate the Riemann tensors of $V_4$ and of the hypersurface in
order to generalize the Gauss and Codazzi equations. Let us first recall that
the definition of the curvature tensor of any connection is \cite{HE}
\begin{eqnarray}
R\left(\vec{X},\vec{Y}\right)\vec{Z}=\nabla_{\vec{X}}\nabla_{\vec{Y}}\vec{Z}-
\nabla_{\vec{Y}}\nabla_{\vec{X}}\vec{Z}-\nabla_{[\vec{X},\vec{Y}]}\vec{Z}
\enspace , \label{rim}
\end{eqnarray}
but using repeatedly equation (\ref{dec}) we have
\begin{eqnarray*}  
\nabla_{\vec{X}}\nabla_{\vec{Y}}\vec{Z}=\nabla_{\vec{X}}\left(\nab_{\vec{Y}}
\vec{Z}-
K\left(\vec{Y},\vec{Z}\right)\vec\r\,\right)=\nab_{\vec{X}}\nab_{\vec{Y}}
\vec{Z}-K\left(\vec{X},\nab_{\vec{Y}}\vec{Z}\right)\vec{\r}- 
\nabla_{\vec{X}}\left(K\left(\vec{Y},\vec{Z}\right)\vec\r\,\right), \\
\nabla_{\left[\vec{X},\vec{Y}\right]}\vec{Z}=\nab_{\left[\vec{X},\vec{Y}
\right]}\vec{Z}-K\left(\left[\vec{X},\vec{Y}\right],\vec{Z}\right)\vec\r 
\hspace{43mm} 
\end{eqnarray*}
so that putting all this into formula (\ref{rim}) we find
\begin{eqnarray}
R\left(\vec{X},\vec{Y}\right)\vec{Z}=\R\left(\vec{X},\vec{Y}\right)\vec{Z}
-K\left(\vec{X},\nab_{\vec{Y}}\vec{Z}\right)\vec\r&+& K\left(\vec{Y},
\nab_{\vec{X}}\vec{Z}\right)
\vec\r-\nabla_{\vec{X}}\left(K\left(\vec{Y},\vec{Z}\right)\vec\r\,\right)+
\nonumber\\
+\nabla_{\vec{Y}}
\left(K\left(\vec{X},\vec{Z}\right)\vec\r\right)&+&K\left(\left[
\vec{X},\vec{Y}\right],\vec{Z}\right)\vec\r \enspace .
\label{Riem}
\end{eqnarray}
Contracting this expression with an arbitrary 1-form $\mbox{\boldmath $\alpha$}
\in A^{\vec\r}$, so that $\mbox{\boldmath$\alpha$}\left(\vec\r\,\right) =0$, we
find the desired generalization of Gauss equation
\begin{eqnarray}
\mbox{\boldmath
$\alpha$}\left(R\left(\vec{X},\vec{Y}\right)\vec{Z}\right)=\mbox{\boldmath
$\alpha$}\left(\R\left(\vec{X},
\vec{Y}\right)\vec{Z}\right) -K\left(\vec{Y},\vec{Z}\right)\mbox{\boldmath
$\alpha$}\left(\nabla_{\vec{X}}
\vec\r\,\right) +K\left(\vec{X},\vec{Z}\right)\mbox{\boldmath
$\alpha$}\left(\nabla_{\vec{Y}}\vec\r\,\right) \label{glob}
\end{eqnarray}
where there appears, in a natural way, a new 1-contravariant, 1-covariant
tensor on $\Sigma$ defined by
$\Psi\left(\mbox{\boldmath$\alpha$},\vec{X}\right)\equiv
\mbox{\boldmath$\alpha$}
\left(\nabla_{\vec{X}}\vec\r\,\right)$ or, in components \cite{S}
\begin{eqnarray}
\Psi_b^a=\omega^a_{\mu} e_b^{\nu}\nabla_{\nu}\r^{\mu} \label{Psi}
\end{eqnarray}
and thus, Gauss' equation takes the form
\begin{eqnarray}
{\omega}^d_{\alpha}R^{\alpha}_{\beta\gamma\delta}e^{\beta}_ae^{\gamma}_b
e^{\delta}_c ={\R}^d_{abc}-K_{ac}{\Psi}^d_b+K_{ab}{\Psi}^d_c \enspace .
\label{Gauss}
\end{eqnarray}
Analogously, contracting (\ref{Riem}) with the normal form $\n$ we get
\begin{eqnarray*}
\n\left(R\left(\vec{X},\vec{Y}\right)\vec{Z}\right)=K\left(\vec{Y},
\nab_{\vec{X}}\vec{Z}\right)-K\left(\vec{X},
\nab_{\vec{Y}}\vec{Z}\right)+K\left(\left[\vec{X},\vec{Y}\right],\vec{Z}
\right)+ \hspace{1cm}\nonumber \\
\nabla_{\vec{Y}}\left(K\left(\vec{X},\vec{Z}\right)\right)-\nabla_{\vec{X}}
\left(K\left(\vec{Y},\vec{Z}\right)\right)-
K\left(\vec{Y},\vec{Z}\right)\n\left(\nabla_{\vec{X}}\vec\r\,\right)+K\left(
\vec{X},\vec{Z}\right)\n
\left(\nabla_{\vec{Y}}\vec\r\,\right) \hspace{2mm}
\end{eqnarray*}
which can be simplified to the following expression, that we call {\it
Codazzi-1 equation},
\begin{eqnarray}
n_{\mu}R^{\mu}_{\alpha\beta\gamma}e_a^{\alpha}e_b^{\beta}e_c^{\gamma}=
\nab_cK_{ba}-\nab_bK_{ca}+K_{ba}\varphi_c-K_{ca}\varphi_b \label{Cod-1}
\end{eqnarray} 
where again a one-form in $\Sigma$ arises naturally: $\mbox{\boldmath
$\varphi$}\left(\vec{X}\right)\equiv{\n}
\left(\nabla_{\vec{X}}\vec\r\,\right)$. Its components are
\begin{eqnarray}
\varphi_a=n_{\mu}e_a^{\nu}\nabla_{\nu}\r^{\mu} \enspace .\label{phi}
\end{eqnarray}
From the Ricci identity $\left(\nabla_{{\vec{e}}_a}\nabla_{{\vec{e}}_b}-
\nabla_{{\vec{e}}_b}\nabla_{{\vec{e}}_a}\right)\r^{\mu}=
R^{\mu}_{\alpha\beta\gamma}\r^{\alpha}e_a^{\beta}e_b^{\gamma}$ and contracting
first with the three forms $\mbox{\boldmath$\omega$}^c$ and second with the
normal form $\n$ one can easily find the following equations
\begin{eqnarray}
\omega^c_{\mu}R^{\mu}_{\alpha\beta\gamma}\r^{\alpha}e_a^{\beta}e_b^{\gamma}
=\nab_a\Psi_b^c-\nab_b\Psi^c_a+\varphi_b\Psi^c_a-\varphi_a\Psi^c_b \enspace ,
\label{Cod-2}\\
n_{\mu}R^{\mu}_{\alpha\beta\gamma}\r^{\alpha}e_a^{\beta}e_b^{\gamma}
=\nab_a\varphi_b-\nab_b\varphi_a+K_{cb}\Psi^c_a-K_{ca}\Psi^c_b \label{Cod-3}
\end{eqnarray}
which we call {\it Codazzi-2 and Codazzi-3 equations}, respectively.

We shall also write down another equation involving the one-form {\boldmath
$\r$} which is obtained from the Ricci identity applied to that one-form and
contracting later with the tangent vectors ${\vec{e}}_c$. The equation reads
\begin{eqnarray}
\r_{\mu}R^{\mu}_{\alpha\beta\gamma}e_a^{\alpha}e_b^{\beta}e_c^{\gamma}=
\nab_c {\cal H}_{ab}- \nab_b{\cal H}_{ac}+ \frac{1}{2} \partial_b \left (
\vec{\r}\cdot \vec{\r} \, \right ) K_{ac}-\frac{1}{2} \partial_c \left (
\vec{\r}\cdot\vec{\r} \, \right ) K_{ab} \label{Cod-4}
\end{eqnarray}
and obviously it is not independent of Codazzi-1 and Gauss equations.  There
appears, however, a new tensor in the hypersurface, in general not symmetric,
defined as
\begin{eqnarray}
{\cal H}_{ab}=e_a^{\mu}e_b^{\nu}\nabla_{\mu} \r_{\nu} \label{Eich}
\end{eqnarray}
which will play a central role in the discussion of the junction conditions in
Sect.5. Codazzi's equations (\ref{Cod-1},\ref{Cod-2}) and expression
(\ref{Cod-4}) collapse into the unique standard Codazzi equation in the case of
non-null hypersurfaces everywhere when the rigging is chosen in the usual way
as the normal vector to the hypersurface. The third Codazzi equation vanishes
identically in that case.

With the definitions we already have in $\Sigma$, we can easily derive the
following formulas for the directional derivatives of different objects along
the vectors ${\vec{e}}_a$:
\begin{eqnarray}
& &\nabla_{{\vec{e}}_a}{\vec{e}}_b=-K_{ab}\vec{\r}+\Gamma^c_{ab} {\vec{e}}_c
\enspace ,\label{rel-1}  \\ & &\nabla_{{\vec{e}}_a}{\n}=-\varphi_a \n+K_{ab}
\mbox{\boldmath$\omega$}^b \enspace ,\label{rel-2} \\
& &\nabla_{{\vec{e}}_a}\vec{ \r}=\varphi_a\vec{\r}+\Psi^b_a {\vec{e}}_b
\enspace ,\label{rel-3} \\ &
&\nabla_{{\vec{e}}_a}\mbox{\boldmath$\omega$}^b=-\Psi^b_a \n-\Gamma^b_{ac}
\mbox{\boldmath$\omega$}^c \enspace .\label{rel-4}
\end{eqnarray}
We established in the last section a way to relate tensors in the manifold
$V_4$ with tensors in $\Sigma$. Now, by means of the previous equations, we are
going to establish a general relation between the covariant derivatives in the
manifold and in the hypersurface. To that aim, let
$S^{\mu_1\cdots\mu_r}_{\nu_1\cdots\nu_q}$ be a tensor field in $V_4$ defined at
least on every point of $\Sigma$. By means of the pull-back and the map $T$,
one can assign to this tensor another tensor field, defined in the
hypersurface, with components
\begin{eqnarray}
S^{a_1\cdots a_r}_{b_1\cdots b_q}={\omega}^{a_1}_{\mu_1}\cdots
{\omega}^{a_r}_{\mu_r}e^{\nu_1}_{b_1}\cdots e^{\nu_q}_{b_q}
S^{\mu_1\cdots\mu_r}_{\nu_1\cdots\nu_q} \enspace .\label{assoc}
\end{eqnarray}
Then, it is not difficult to prove that the projection to the hypersurface of
the covariant derivative of $S^{\mu_1\cdots\mu_r}_{\nu_1\cdots\nu_q}$ is
related to the covariant derivative of $S^{a_1\cdots a_r}_{b_1\cdots b_q}$ in
the following form
\begin{eqnarray}
{\omega}^{a_1}_{\mu_1}\cdots {\omega}^{a_r}_{\mu_r}e^{\nu_1}_{b_1}\cdots
e^{\nu_q}_{b_q}e^{\gamma}_c
\nabla_{\gamma}S^{\mu_1\cdots\mu_r}_{\nu_1\cdots\nu_q}=\nab_c
S^{a_1\cdots a_r}_{b_1\cdots b_q}+\sum_{i=1}^{r} S^{a_1\cdots a_{i-1}\gamma
a_{i+1} \cdots a_r}_{b_1\cdots b_q}n_{\gamma}
\Psi^{a_i}_c +\nonumber \\
+\sum_{j=1}^{q}S^{a_1\cdots a_r}_{b_1\cdots b_{j-1} \gamma b_{j+1}
\cdots b_q}\r^{\gamma}K_{cb_i} \enspace ,\hspace{3.5cm} \label{deri}
\end{eqnarray}
where the mixed tensor $S^{a_1\cdots a_{i-1}\gamma a_{i+1} \cdots
a_r}_{b_1\cdots b_q}$ is the projection of the original tensor in the surface
on all its indexes except $\gamma$ and analogously for $S^{a_1\cdots
a_r}_{b_1\cdots b_{j-1}
\gamma b_{j+1}\cdots b_q}$. Immediate consequences of this equation are
\begin{eqnarray}
\nab_c\g_{ab}+\r_bK_{ca}+\r_aK_{cb}=0 \enspace , \hspace{2.2cm} \label{ge} \\
\nab_c\r_a=-\left( \vec{\r}\cdot\vec{\r} \,\right ) K_{ca}+\g_{ab}\Psi^b_c
+\r_a\varphi_c =-\left( \vec{\r}\cdot\vec{\r} \,\right ) K_{ca}+ {\cal H}_{ca}
\enspace .\label{derl}
\end{eqnarray}

From the Riemann tensor of the rigged connection we can define, following the
usual notation of the Ricci calculus \cite{S}, the tensor fields
\begin{eqnarray}
& &\R_{bd}=\R^a_{bad} \hspace{2cm}{\rm (Ricci \ tensor),} \label{Ricc} \\ &
&V_{cd}=\R^a_{acd} \enspace .\label{V}
\end{eqnarray}
This last tensor is not identically zero because the connection in the
hypersurface is not metric in general. Keeping in mind that the torsion of the
connection vanishes, these tensors just defined and the Riemann tensor satisfy
the following identities
\begin{enumerate}
\item  $\R^a_{bcd}=-\R^a_{bdc}\, .$
\item  $\R^a_{bcd}+\R^a_{cdb}+\R^a_{dbc}=0 \ \ \ $ (First Bianchi identity).
\item  $V_{cd}=-V_{dc}\, .$
\item  $V_{cd}=\R_{cd}-\R_{dc}\, .$
\item  $\nab_e\R^a_{bcd}+\nab_c\R^a_{bde}+\nab_d\R^a_{bec}=0 \ \ \ $ (Second
Bianchi identity).
\item  $\nab_aV_{bc}+\nab_bV_{ca}+\nab_cV_{ab}=0 \ \Leftrightarrow \ {\rm d}
{\bf V}=0 \, .$
\end{enumerate}
Last identity shows that the two-form ${\bf V}$ is closed and then, because of
the Poincar\'e Lemma, it is locally exact. Actually it is globally exact as the
following result proves:
\begin{proposition}
The two-form ${\bf V}$ is, in fact, exact and it is related with
{\boldmath$\varphi$} defined above by ${\bf V}=-{\rm
d}\mbox{\boldmath$\varphi$}$ .
\end{proposition}
{\it Proof: } The proof we present here of this result makes a main use of the
Gauss and Codazzi-3 equations. Contracting the first two indexes in the Gauss
equation, we find
$V_{bc}=\R^a_{abc}=K_{ac}\Psi^a_b-K_{ab}\Psi^a_c+R^{\mu}_{\alpha\beta\gamma}
e^{\alpha}_a e^{\beta}_b e^{\gamma}_c \omega^a_{\mu}$ and using now formula
(\ref{proj}) we obtain
$V_{bc}=K_{ac}\Psi^a_b-K_{ab}\Psi^a_c-n_{\mu}R^{\mu}_{\alpha\beta\gamma}
\r^{\alpha}e_b^{\beta}e_c^{\gamma}$. Codazzi-3 equation leads us then to 
\begin{eqnarray*} 
V_{bc}=-\left(\partial_b\varphi_c-\partial_c\varphi_b\right),
\end{eqnarray*}
where we have used that the rigged connection is torsion-free, and the proof is
complete.

\vspace{1cm}

On the other hand, the expression of the Riemann tensor in terms of the
Christoffel symbols 
${\bf V}$ in terms of them:
\begin{eqnarray*}
V_{bc}=\partial_b\Gamma^a_{ac}-\partial_c\Gamma^a_{ab} \enspace ,
\end{eqnarray*}
and this together with the previous expression involving $V_{bc}$ suggests the
definition of the geometric object
\begin{eqnarray*}
\Gamma_c \equiv \varphi_c+\Gamma^a_{ac}
\end{eqnarray*}
which verifies $\partial_b\Gamma_c-\partial_c\Gamma_b=0$.  We cannot say that
$\Gamma_c$ is a closed 1-form because $\Gamma_c$ is not a tensor. However, in
each coordinate system of the hypersurface there is a function $\phi$ such
that, locally
\begin{eqnarray*}
\Gamma_c=\partial_c \phi \enspace .
\end{eqnarray*}

Until now we have not considered the transformation of the above defined
objects by choosing the rigging in another direction. Let us change the
direction of the rigging without changing the normalization of the normal form
and maintaining the condition $n_{\mu}\r^{\mu}=1$ everywhere. Under these
assumptions, the most general change of the rigging field is
\begin{eqnarray}
\r^{'\mu}=\r^{\mu}+s^{\mu} \label{camb}
\end{eqnarray}
where the vector $s^{\mu}$ verifies $s^{\mu}n_{\mu}=0$ and therefore
$s^{\mu}=s^a e_a^{\mu}$. This change of the rigging induces the following
transformation on the objects depending on it
\begin{eqnarray}
& &\mbox{\boldmath$\omega'$}^a=\mbox{\boldmath$\omega$}^a-s^a \n ,
\nonumber \\ 
& &\Gamma'^a_{bc}=\Gamma^a_{bc}+s^aK_{bc}\, , \nonumber \\ &
&\varphi'_a=\varphi_a-s^bK_{ba} \, ,  \label{Camb}\\ &
&\Psi'^a_b=\Psi^a_b-s^a\varphi_b+s^a s^c K_{bc}+\nab_bs^a
\nonumber
\end{eqnarray}
and from these expressions we immediately obtain the following interesting
result.
\begin{proposition}
The object $\Gamma_c$ does not depend on the rigging.
\end{proposition}
In consequence, the function $\phi$ related with $\Gamma_c$ as was explained
above does not depend on the direction of the rigging. Let us find now how the
object $\Gamma_c$ transforms under a change of coordinates
$\xi^a=\xi^a\left(\xi'^a\right)$ in $\Sigma$. If we call
$A^a_{a'}=\frac{\partial\xi^a}{\partial\xi'^a}$ the jacobian matrix of this
transformation, the transformation law for $\Gamma_c$ is
\begin{eqnarray*}
\Gamma_{a'}=A^a_{a'}\Gamma_a+\frac{1}{\det\left(A\right)}
\partial_{a'}\det\left(A\right)
\end{eqnarray*}
Using now that in both coordinate systems $\Gamma_c$ is the partial derivative
of a function $\phi$, we can easily relate the function in a coordinate system
with the function in the other one by $\phi'=\phi+\log\left(\left\vert
\det\left(A\right)\right\vert\right)$, or equivalently, $e^{\phi'}=\left\vert
\det\left( A\right)\right\vert e^{\phi}$.  Thus, for orientation preserving
coordinate changes, $e^{\phi}$ transforms as a scalar density of weight +1 and
therefore, it may have something to do with the volume element in the
hypersurface. In fact, this relation is concreted as follows.
\begin{proposition}
Except for a multiplicative constant, the function $e^{\phi}$ is the unique
independent component of the volume form already defined in the hypersurface.
\end{proposition}
{\it Proof: } From definition (\ref{vol}) and using the normalization condition
$\r^{\mu} n_{\mu}=1$ we have
$\eta_{123}=\r^{\alpha}e_1^{\beta}e_2^{\gamma}e_3^{\delta}
\eta_{\alpha\beta\gamma\delta}$, expression that, despite of its appearance,
does not depend on the specific choice of $\vec\r$. If we calculate now
\begin{eqnarray*}
\partial_a\left(\eta_{123}\right)=e_a^{\rho}\nabla_{\rho}\left(\r^{\alpha}
e_1^{\beta}
e_2^{\gamma}e_3^{\gamma}\eta_{\alpha\beta\gamma\delta}\right)=\eta_{\alpha\beta
\gamma\delta} e_a^{\rho}\nabla_{\rho}\left(\r^{\alpha}e_1^{\beta}e_2^{\gamma}
e_3^{\gamma}\right)
\end{eqnarray*}
and make use of formulas (\ref{rel-1}), (\ref{rel-3}) and the complete
antisymmetry of $\eta_{\alpha\beta\gamma\delta}$, we get
\begin{eqnarray}
\partial_a\eta_{123}=\eta_{\alpha\beta\gamma\delta}\left[\varphi_a\r^{\alpha}
e_1^{\beta}e_2^{\gamma}e_3^{\delta}+\Gamma^1_{a1}\r^{\alpha}
e_1^{\beta}e_2^{\gamma}e_3^{\delta}+\Gamma^2_{a2}\r^{\alpha}
e_1^{\beta}e_2^{\gamma}e_3^{\delta}+ \right.\nonumber \\
\left.\Gamma^3_{a3}\r^{\alpha}e_1^{\beta}e_2^{\gamma}e_3^{\delta}\right]
=\eta_{123}\left(\varphi_a+\Gamma^c_{ac}\right)=\eta_{123}\Gamma_a &
\label{fi}
\end{eqnarray}
so that being $\eta_{123}$ positive everywhere we find $\eta_{123}=Ce^{\phi}$
where $C$ is a positive constant.

\vspace{1cm}

Let us change now the normalization factor of the normal form, without changing
the direction of the rigging vector $\vec\r$ anywhere. So, we put
$n'_{\mu}=\sigma n_{\mu}$ and if we want to keep the volume form positive, the
factor $\sigma$ must be positive everywhere. The changes induced by this
transformation are
\begin{eqnarray}
\eta'_{123}=\frac{1}{\sigma} \eta_{123}, \hspace{4mm} K'_{ab}=\sigma K_{ab}, 
\hspace{4mm} \vec{\r'}=\frac{1}{\sigma}\vec{\r},\hspace{4mm} 
{\Psi'}^a_b=\frac{1}{\sigma}\Psi^a_b \nonumber \\
\mbox{\boldmath$\omega$}^a_{\mu}{\rm \
and \ }\Gamma^a_{bc} {\rm \ remain \ invariant,  } \hspace{2mm}
\varphi'_a=\varphi_a - \partial_a\log\sigma , \hspace{2mm}
\Gamma'_a=\Gamma_a-\partial_a\log\sigma . \label{cambs} 
\end{eqnarray}

It is a well-known fact in differential geometry that in an affine manifold
(that is, with a linear connection) possessing a well-defined volume form
$\mbox{\boldmath$\eta$}$, the Stockes theorem can be rewritten as a Gauss
theorem
\[ \int_{U}\nabla_{\mu}X^{\mu}\mbox{\boldmath$\eta$}=
\int_{\partial U}X^{\mu}d\sigma_{\mu}, \]
if and only if the connection is volume preserving, that is to say $\nabla
\mbox{\boldmath$\eta$} = 0$. In the previous formula $U$ is an open
neighbourhood in the manifold, $X^{\mu}$ is any vector field defined at least
on $U$ and $d\sigma_{\mu}$ is the normal volume form defined on the boundary
$\partial U$ of $U$. Of course, for all metric connections this is immediately
true and the Gauss theorem holds in the spacetime.  In our case, however, we
have defined in the hypersurface a volume form and a connection that are not
metric and then we are not sure that the Gauss theorem holds in the
hypersurface. Therefore, we want to ascertain under which conditions the rigged
connections in the hypersurface are volume preserving. To that aim, let us
calculate
\begin{eqnarray*}
\nab_a\eta_{bcd}=\partial_a\eta_{bcd}-\Gamma^{e}_{ba}\eta_{ecd}-\Gamma^{e}_{ca}
\eta_{bed}-\Gamma^{e}_{da}\eta_{bce}=\partial_a\eta_{bcd}-\Gamma^e_{ea}
\eta_{bcd}
\end{eqnarray*}
where we have used the antisymmetry property of $\mbox{\boldmath$\eta$}$.
Making use here of formula (\ref{fi}) we can write then
\begin{eqnarray}
\nab_a\eta_{bcd}=\varphi_a\eta_{bcd} \enspace . \label{cov}
\end{eqnarray}
In the same way, it is easy to find the analogous expression for the
contravariant volume form:
\begin{eqnarray}
\nab_a\eta^{bcd}=-\varphi_a\eta^{bcd} \enspace . \label{cont}
\end{eqnarray} 
Consequently, we have proven the following standard result.
\begin{theorem}
Given a normal form and a rigging vector, the necessary and sufficient
condition for the rigged connection to be volume preserving is that
$\mbox{\boldmath $\varphi$}=0$.
\end{theorem}
The condition $\mbox{\boldmath $\varphi$}=0$ is called by Schouten \cite{S} the
{\it second condition of normalization}.

Looking now at the transformation formula (\ref{cambs}) for
$\mbox{\boldmath$\varphi$}$, we see that the above theorem can be improved if
we allow for changes in the normalization of the normal form (or the volume
element). In this sense, it is immediate the following theorem.
\begin{theorem}
Given a rigging direction, the necessary and sufficient condition such that
there exists a normalization of the normal form (or equivalently a volume form
in the hypersurface) in which the rigged connection is volume preserving is
that the 1-form $\mbox{\boldmath $\varphi$}$ be exact. Moreover, this volume
form, if it exists, is unique except for a constant factor.
\end{theorem}
The question of whether or not there always exists a rigging and a
normalization factor of the normal form such that the rigged connection is
volume preserving is a little bit more complicated. In the case where the
second fundamental form is not degenerate (or even degenerate only once), the
existence of such a connection can be shown without difficulty. In other cases,
however, that depends on some integrability conditions involving both the
manifold and hypersurface structures. In the next section we present another
type of connection in the hypersurface which is always volume preserving.
\section{Second Connection in a General Hypersurface: The Rigged Metric
Connection.} We are now going to develop a completely different method to
define a connection in the hypersurface. As before, we begin with a rigging
vector field $\vec\r$ defined on every point in $\Sigma$ which verifies
$\n\left(\vec\r\,\right)=1$.  We have already studied $\Phi^{\star}(g)$ and we
have shown that the first fundamental form is degenerate if and only if the
normal form is null.  We consider now the symmetric non-degenerate tensor
$g^{\mu\nu}$ and its projection into the hypersurface by means of the map $T$.
Its components are
\[ g^{ab}=g^{\mu\nu}\omega^a_{\mu}\omega^b_{\nu} \]
and obviously they do depend on the rigging of the hypersurface.  With this new
tensor in the hypersurface we can complete relations (\ref{var}) with the
following formulas
\begin{eqnarray}
& g^{ab}\r_b=-\left( \vec{\r}\cdot\vec{\r} \,\right )n^a \, ,\hspace{5mm}&
g^{ab}\g_{bc}=\delta^a_c -n^a\r_c \enspace .\label{mes}
\end{eqnarray}
Let us find a necessary and sufficient condition for this symmetric
contravariant tensor to be non-degenerate. First of all we establish the
following lemma.
\begin{lemma}
The rigging vector $\vec\r$ can be expressed in the following form
\[ \r^{\mu}=-\eta_{123}\eta^{\mu\nu\rho\lambda}\omega^1_{\nu}\omega^2_{\rho}
\omega^3_{\lambda} \]
\end{lemma}
{\it Proof: } The proportionality between $\r^{\mu}$ and the vector
$\chi^{\mu}\equiv\eta^{\mu\nu\rho\lambda}\omega^1_{\nu}
\omega^2_{\rho}\omega^3_{\lambda}$ follows from the relation 
$\omega^a_{\mu}\chi^{\mu}=0$ for every $a$, so we can write
$\r^{\mu}=B\eta^{\mu\nu\rho\lambda}\omega^1_{\nu}\omega^2_{\rho}
\omega^3_{\lambda} $ for some factor $B$. Using now that $\r^{\mu}n_{\mu}=1$
and formula (\ref{ene}) we have
\[ 1=n_{\mu}\r^{\mu}=\frac{B}{\eta_{123}}\eta_{\mu\beta\gamma\delta}
\eta^{\mu\nu\rho\lambda}e_1^{\beta}e_2^{\gamma}e_3^{\delta}\omega^1_{\nu}
\omega^2_{\rho}\omega^3_{\lambda}=-\frac{B}{\eta_{123}} \]
and therefore $B=-\eta_{123}$\, , as we wanted to prove.

\vspace{1cm}

An easy calculation using the above expression for $\vec\r$ shows the following
result
\begin{eqnarray}
\r^{\alpha}\r_{\alpha}=- {\left(\eta_{123}\right)}^2 \det\left(
g^{ab} \right ) \label{ele}
\end{eqnarray}
which is similar to formula (\ref{nor}), and then, analogously to Lemma 1, we
have:
\begin{proposition}
$g^{ab}$ is degenerate in a point $x \in \Sigma$ if and only if the rigging
vector is null at that point.
\end{proposition}
There is, though, an important difference between the uselfulness of this result
and that of the similar one for the normal form $\n$. Whereas the normal form
$\n$ is determined by the hypersurface, the rigging vector $\vec{\r}$ can be
chosen in many different ways and, in particular, it can be taken non-null
everywhere in $\Sigma$ so that the tensor $g^{ab}$ is non-degenerate everywhere
in the hypersurface. In this case, we have that the tensor $g^{ab}$ has an
inverse, that we will call $\gamma_{ab}$, satisfying
$g^{ab}\gamma_{bc}=\delta^a_c$. It is easy to see that this tensor can be
written
\begin{eqnarray}
\gamma_{ab}=\g_{ab}-\frac{1}{\left (\vec{\r}\cdot\vec{\r}\,\right )}\r_a\r_b
\enspace .\label{metr}
\end{eqnarray}
Of course, there are many symmetric two-covariant tensors in the manifold such
that $\Phi^{\star}$ applied to them gives $\gamma_{ab}$, but we have already
defined a privileged one among them, which we will call $\gamma_{\mu\nu}$, by
demanding that $\r^{\mu}\gamma_{\mu\nu}=0$ or, in other words, by applying
$\Lambda$ to $\gamma_{ab}$. This tensor is
\begin{eqnarray}
\gamma_{\mu\nu}=g_{\mu\nu}-\frac{1}{\left (\vec{\r}\cdot\vec{\r}\,\right )}
\r_{\mu}\r_{\nu} \label{metva}
\end{eqnarray}
from where we can guess a possible physical interpretation of this
construction.  The form of the tensor (\ref{metva}) suggests that we can take
the rigging vector as a timelike vector in the manifold, or equivalently, as an
{\it observer}, and then we define for every hypersurface in the manifold a
non-degenerate two-covariant symmetric tensor by restricting the metric in the
canonical three-spaces of the observer to the hypersurface. The freedom in
choosing the rigging is simply then the possibility of changing the observer in
spacetime.  Thus, different riggings induce different tensors of type
(\ref{metr}) in a given hypersurface, but this would just represent the fact
that different observers would ``see'' the same hypersurface with different
metric properties.  This possibility was in fact suggested some years ago, in a
somewhat different form and for a very particular case, in \cite{B}.

From a mathematical point of view, however, the important thing is that we have
defined a symmetric two-covariant non-degenerate tensor in the hypersurface,
and thereby we have the unique metric connection associated to it, given by
\begin{eqnarray}
\Gam^a_{bc}=\frac{1}{2}g^{ad}\left(\partial_b\gamma_{dc}+\partial_c\gamma_{db}-
\partial_d\gamma_{bc}\right) \label{connex}
\end{eqnarray}
that we call the {\it rigged metric connection}.

Let us now try to relate the rigged connection and the rigged metric connection
defined by the same (non-null) rigging.  Because of the relation
$\nab_a\gamma_{bc}=\partial_a\gamma_{bc}
-\Gamma^e_{ba}\gamma_{ec}-\Gamma^e_{ca}\gamma_{ae}$, we can rewrite the formula
defining $\Gam^a_{bc}$ as
\begin{eqnarray}
\Gam^a_{bc}=\frac{1}{2}g^{ad}\left(\nab_b\gamma_{dc}+\nab_c\gamma_{db}-
\nab_d\gamma_{bc} \right)+ \Gamma^a_{bc}  \label{connexa}
\end{eqnarray}
which makes explicit the following obvious statement:
\begin{proposition}
Given a rigging, the two connections $\Gam$ and $\Gamma$ coincide iff
$\nab_{d}\gamma_{bc}=0$.
\end{proposition}
We are going to reexpress this result in a more interesting and useful form by
using expression (\ref{metr}) and expanding $\nab_a\gamma_{bc}$.  We only need
to know the covariant derivatives of $\g_{ab}$ and $l_c$ (formulas (\ref{ge})
and (\ref{derl})\,), as well as the differential of the norm of the rigging,
$\partial_c \left(\vec{\r}\cdot\vec{\r} \,\right )$.  This can be found
staightforwardly from formula (\ref{rel-3})
\begin{eqnarray}
\partial_c\left( \vec{\r}\cdot\vec{\r}\,\right )= 2\left (\vec{\r}\cdot\vec{\r}
\,\right )
\varphi _c +2\Psi^e_c\r_e \label{derpr}
\end{eqnarray}
and putting all this together we easily find
\begin{eqnarray*}
\nab_d\gamma_{bc}=-\frac{1}{\left( \vec{\r}\cdot\vec{\r} \,\right )} \Psi^e_d
\left(
\gamma_{ec}\r_b+\gamma_{eb}\r_c \right ) \enspace .
\end{eqnarray*}
\begin{theorem}
Given a non-null rigging, the necessary and sufficient condition such that the
rigged connection $\Gamma$ and the rigged metric connection $\Gam$ coincide is
that, for any point $p \in \Sigma$, either $\left .\Psi^a_b \right \vert_p=0$
or \hspace{2mm} $\r_a \mid_p=0$.
\end{theorem}
{\it Proof:} The necessary and sufficient condition is that
$\nab_{a}\gamma_{bc}=0$ or, what is the same,
$\Psi^e_c\left(\gamma_{ea}\r_b+\gamma_{eb}\r_a \right )=0$. But we know that if
an object of type $U_aV_b+U_bV_a$ vanishes then necessarily $U_a=0$ or $V_a=0$
.  Applying this result to the previous expression, and noting that the index
$c$ plays no role in this reasoning, we have $\Psi^e_c\gamma_{ea}=0$ or
$\r_a=0$, and as the metric $\gamma_{ab}$ is not degenerate the theorem
follows.

\vspace{0.5cm}

From its definition, it is trivial that $\r_c=0$ if and only if $\vec{\r}$ is
proportional to the vector $\vec{n}$ ($\r_a=0 \Leftrightarrow
\r_{\mu}e_a^{\mu}=0 \ \forall a \Leftrightarrow \r_{\alpha} \ {\rm \
is \ proportional \ to \ } n_{\alpha} $).  In the case of hypersurfaces
non-null everywhere we can choose the normal vector $\vec{n}$ as the rigging
vector and then we have $\r_a=n_a=0$.  With this choice of the rigging the two
connections coincide, as is obvious from its construction, giving the natural
connection in the hypersurface. In the general case, however, if there is some
point where the hypersurface is null, we cannot take the normal vector as the
rigging and then $\r_a\neq 0$ for some $a$. So, in the general case, the only
possibility for having a rigging such that the two connections coincide is that
$\Psi^a_b=0$.  In consequence, we are now going to study under which conditions
one can choose a rigging such that $\Psi^a_c=0$ everywhere in the hypersurface.
Recalling expression (\ref{rel-3}) we have that $\Psi^a_b=0$ is equivalent to
\begin{eqnarray}
e_b^{\nu}\nabla_{\nu}\r^{\mu}=\varphi_b\r^{\mu} \label{P}
\end{eqnarray}
and then contracting with vector $\vec{\r}$ we find
\begin{eqnarray*}
\partial_b \left( \log \left( \sqrt { \left \vert \vec{\r}\cdot\vec{\r} 
\right \vert} \right) \right )= \varphi_b 
\end{eqnarray*}
so that a necessary condition for the equation (\ref{P}) to hold is that
{\boldmath $\varphi$} be an exact 1-form. We can now define a new vector field
\begin{eqnarray*}
\vec{V} \equiv \frac{\vec{\r}}{\sqrt{ \left \vert \vec{\r}\cdot\vec{\r}
\,\right
\vert }}
\end{eqnarray*}
and it is very easy to check that if the vector $\vec{\r}$ verifies equation
(\ref{P}), then the vector $\vec{V}$ verifies
\begin{eqnarray} 
e_b^{\nu}\nabla_{\nu}V^{\mu}=0 \label{P2}
\end{eqnarray}
and conversely, if $\vec{V}$ verifies equation (\ref{P2}) then any vector of
the form $\vec{\r} \equiv \sigma \vec{V}$ verifies equation (\ref{P}), where
$\sigma$ is any non-vanishing function in the hypersurface.  This last equation
is written in normal form, and then we can study its integrability conditions
by the standard procedure. These integrability conditions are found to be
\begin{eqnarray}
R^{\mu}_{\sigma\alpha\beta}e_c^{\alpha}e_b^{\beta}V^{\sigma}=0 \label{intr-1}
\end{eqnarray}
When this relation is identically satisfied, that is to say when
$R^{\mu}_{\sigma\alpha\beta}e_c^{\alpha}e_b^{\beta}=0$, then equation
(\ref{P2}) has a general solution depending on three constant parameters (in
principle there are four constants that are the initial conditions $\left.
V^{\mu}\right \vert _p$ at some point $p \in \Sigma$, but they are subject to
the relation $\left.  V^{\mu}\cdot V_{\mu}\right \vert_p=\pm1$). If they are
not identically verified, we must go on with the integrability conditions,
which are successively
\begin{eqnarray}
e_d^{\rho}\nabla_{\rho}\left(R^{\mu}_{\sigma\alpha\beta}e_c^{\alpha}e_b^{\beta}
\right)V^{\sigma}=0 \label{intr-2} \\
e_f^{\kappa}\nabla_{\kappa}\left(e_d^{\rho}\nabla_{\rho}\left(R^{\mu}_{\sigma
\alpha\beta}e_c^{\alpha}e_b^{\beta}\right)\right)V^{\sigma}=0 \label{intr-3} \\
e_g^{\lambda}\nabla_{\lambda}\left(e_f^{\kappa}\nabla_{\kappa}\left(e_d^{\rho}
\nabla_{\rho}\left(R^{\mu}_{\sigma
\alpha\beta}e_c^{\alpha}e_b^{\beta}\right)\right)\right)V^{\sigma}=0
\label{intr-4} \\
\vdots \nonumber
\end{eqnarray}
If the first equation in the row is verified identically (and the previous
integrability condition is not) then the general solution has as many arbitrary
constants of integration as the number of solutions (if any) for $V^{\alpha}$
of the equation (\ref{intr-1}) at any point in the hypersurface. The following
integration conditions must be understood in a similar form. It is evident that
the existence or not of a rigging whose two connections coincide depends both
on the form of the hypersurface and on the manifold in which it is imbedded.

To end this section, we shall establish the relation between $\Gam^a_{ac}$ and
$\Gamma^a_{ac}$. Recalling formula (\ref{ele}): $\left( \vec{\r}\cdot\vec{\r}
\,\right ) = - {\left ( \eta_{123} \right )}^2 \det \left ( g^{ab} \right ) = -
\frac{{\left ( \eta_{123} \right )}^2}{ \det \left ( \gamma_{ab} \right ) }$
and using the fact that $\Gam^a_{ac}= \partial_c \left (\log \left (\sqrt
{\left \vert \det \left ( \gamma_{ab} \right ) \right \vert } \right )\right )$
we find
\begin{eqnarray}
\Gam^a_{ac}= \Gamma^a_{ac} + \varphi_c - \frac{1}{2}\partial_c \left (\log 
\left (\vert \vec{\r}\cdot\vec{\r} \, \vert  \right )\right )=
\Gamma_c - \frac{1}{2}\partial_c \left ( \log \left (  \vert \vec{\r}\cdot
\vec{\r} \, \vert  \right )\right )
\label{expr2}
\end{eqnarray}
which is the desired relation. These formulas lead us to the following
proposition.
\begin{proposition}
The volume preserving connection $\Gam$ is such that $\Gam^a_{ac}=\Gamma_c$
when $\left( \vec{\r}\cdot\vec{\r} \,\right )$ is chosen to be constant in the
hypersurface.
\end{proposition}

Therefore, the rigged metric connections have some advantages like being always
volume preserving and the fact that each one fixes a unique volume element in
the hypersurface. They are also good connections from the physical point of
view because, as explained above, they can be interpreted as connections
associated with observers in space-time. Nevertheless, the geometrical meaning
of the rigged metric connections is obscure, contrarily to what happens with
the rigged connections which are defined simply by decomposing the tangent
spaces and then projecting to the hypersurface (with respect to the rigging).
In any case, both constructions provide a geometrical structure to general
hypersurfaces {\it imbedded in spacetime}, and they will also allow us to
define the proper junction conditions in general relativity for arbitrary
hypersurfaces of discontinuity, regardless of its time-, space- or light-like
character, which can also vary from point to point.  This is the subject of the
next section.
\section{Junction Conditions.}
Consider now two $C^3$ spacetimes $V^+$ and $V^-$, each of them with boundary
$\Sigma ^+$ and $\Sigma ^-$ and $C^2$ metrics $g^+$ and $g^-$. Assume further
that there is a $C^3$ diffeomorphism from $\Sigma ^-$ to $\Sigma^+$.  This is
the typical situation in General Relativity where the glueing of two spacetimes
by means of identification of points on the boundaries is considered in order
to study their possible posterior matching. However, as pointed out by Clarke
and Dray \cite{CD}, the mere identification of points in $\Sigma ^+$ and
$\Sigma ^-$ does {\it not} by itself give a well defined geometry in the sense
that one should also specify how the tangent spaces are to be identified. To
clarify this, let us define $V_4$, the whole spacetime, as the disjoint union
of $V^+$ and $V^-$ with diffeomorphically related points in $\Sigma ^+$ and
$\Sigma ^-$ identified. On the complementary of $\Sigma ^{\pm}$ in $V_4$ we
have the metric $g$ given by $g^+$ in $V^+$ and $g^-$ in $V^-$.  The main
result proven by Clarke and Dray reads as follows.
\begin{theorem}
Under the above assumptions, there exists a unique $C^1$ atlas on $V_4$ which
induces the given $C^3$ structures on $V^+$ and $V^-$ and such that $g$ admits
a continuous extension to the whole $V_4$ (and which is maximal with respect to
these properties) if and only if $\,\Sigma ^+$ and $\Sigma ^-$ are isometrical
with respect to their first fundamental forms inherited from $V^+$ and $V^-$;
that is to say, if and only if their respective first fundamental forms
$\g\,^+$ and $\g\,^-$ agree.
\end{theorem}
To be precise, Clarke and Dray proved this theorem under the added assumption
that the signatures of $\g\,^+$ and $\g\,^-$ are constant. This assumption is
superfluous, though, and their proof can be easily generalized to arbitrary
hypersurfaces in which the signature can change from point to point.

The above theorem is of great importance, because if one wishes to define
Einstein's equations, even in the distributional sense, it is necessary that
the metric of the spacetime be, at least, continuous. Thus, if we consider the
whole spacetime $V_4$ and denote simply by $\Sigma$ the image of $\Sigma ^+$ or
$\Sigma ^-$ in it, the necessary and sufficient condition such that the glued
spacetime $V_4$ has well-defined Einstein's equations for the metric $g$ is
that, in a given coordinate system of $\Sigma$, the first fundamental form of
$\Sigma$ calculated from $V^+$ coincide with the first fundamental form of
$\Sigma$ calculated from $V^-$:
\begin{eqnarray}
\g\, ^+ =\g\, ^- \enspace . \label{pjc}
\end{eqnarray}
We call relations (\ref{pjc}) the {\it preliminary junction conditions}. In a
practical problem, these conditions work as follows. We are given two
imbeddings $x^{\mu}_{\pm}=x^{\mu}_{\pm}\left(\xi ^a\right)$ of $\Sigma$, where
$x^{\mu}_{\pm}$ are local coordinates for $V^{\pm}$, respectively, and $\xi^a$
are intrinsic coordinates for $\Sigma$.  Therefore, we also have the vectors
tangent to $\Sigma$: $\vec{e}_a{}^{\pm}$. But if the preliminary junction
conditions hold, then we have in the coordinate system $\{\xi^a\}$
\begin{eqnarray*}
\g_{ab}^+=\g_{ab}^-
\end{eqnarray*}
or, equivalently, the scalar products of the vectors $\vec{e}_a{}^{\pm}$
coincide from $V^+$ and $V^-$. There only remains to choose the riggings
$\vec{\r}{}^{\pm}$ such that $\{\vec{\r}^{\pm},\vec{e}_a{}^{\pm}\}$ are both
bases with the same orientation satisfying
\begin{eqnarray*}
\r_a^+=\r_a^- \, \hspace{1cm} 
\left(\r \cdot \r\right)^+=\left(\r \cdot \r\right)^-
\end{eqnarray*}
and then identify the bases in the tangent spaces
$\{\vec{\r}^+,\vec{e}_a{}^+\}\equiv \{\vec{\r}^-,\vec{e}_a{}^-\}
\equiv \{\vec{\r},\vec{e}_a\}$ {\it by definition} dropping the $\pm$.
In the resulting spacetime, and due to the above theorem, there exists a unique
structure with coordinate systems such that the metric $g$ of $V_4$ is
continuous, hence, the components $\r^{\mu}$ and $e^{\mu}_a$ in these
coordinate systems are well defined.

From now on, we assume that the preliminary junction conditions hold such that
the above construction has been carried out and the whole spacetime $(V_4,g)$
has a hypersurface $\Sigma$ splitting the manifold into the two open sets
$V^{-}$ and $V^{+}$, whose boundary is $\Sigma$, and such that the metric
tensor $g$ is continuous on the whole manifold and at least of type $C^2$ in
both $V^{-}$ and $V^{+}$. We shall also assume that the derivatives up to
second order of this tensor field have a well-defined limit on the hypersurface
of separation $\Sigma$ coming from both $V^{-}$ and $V^{+}$. We will not
restrict the type of the hypersurface in any way whatsoever and, consequently,
we will use the theory of general hypersurfaces developed above. Our aim is to
obtain the curvature tensors of such a spacetime and thereby to find the
necessary and sufficient junction conditions which forbid the existence of
singular parts in the curvature (sometimes these singular parts are called
surface layers or impulsive gravitational waves). The natural way to study this
sort of problems is by using the theory of tensor distributions on manifolds.
For a brief summary and for notations used from here on, see the appendix in
this paper.
 
First of all, we write the step on $\Sigma$ of the derivative of a function
$f$.  It is not difficult to see that, for every vector $\vec{V}$ tangent to
the hypersurface, we have
\begin{eqnarray*}
V^{\mu} \left [ \partial_{\mu} f \right ] = V^{\mu} \partial_{\mu} \left [ f
\right ]
\end{eqnarray*}
and therefore, using the basis $\left \{ \n,{\mbox{\boldmath$\omega$}}^{a}
\right \}$ of the dual tangent plane, we obtain
\begin{eqnarray}
\left [ \partial_{\mu} f \right ]= A n_{\mu}+ \omega^a_{\mu} \partial_a
\left [ f \right ] \label{partderf}
\end{eqnarray}
where $A$ is a scalar function on $\Sigma$ defined by $A\equiv \r^{\mu}
\left [ \partial_{\mu} f \right ]$. Obviously, $A$ depends on the rigging, but
formula (\ref{partderf}) does not. Thus, when the function $f$ is continuous
across $\Sigma$, $A$ is independent of the rigging. This is what happens with
the metric tensor itself, and from the previous equation it follows that
\begin{eqnarray}
\left[ \partial_{\beta}g_{\lambda\rho}\right] = \zeta_{\lambda\rho}n_{\beta}
\label{der1}
\end{eqnarray}
where $\zeta_{\lambda\rho}$ is a two-covariant symmetric tensor field defined
on the hypersurface $\Sigma$ and independent of the rigging. Using this formula
it is immediate to find
\begin{eqnarray}
\left[ \Gamma^{\alpha}_{\beta\gamma} \right ] = \frac{1}{2} \left ( 
\zeta^{\alpha}_{\lambda}n_{\beta}+\zeta^{\alpha}_{\beta} n_{\lambda}
- n^{\alpha}\zeta_{\beta\lambda} \right ) \enspace , \label{Crissym}
\end{eqnarray}
and substituting this in formula (\ref{dRiem}) of the appendix, we get
\begin{eqnarray}
{\underline{R}}^{\alpha}_{\beta\lambda\mu}=\left(\1-\otheta\right)
\cdot{R^{-}}^{\alpha}_{\beta\lambda\mu}
+\otheta\cdot {R^{+}}^{\alpha}_{\beta\lambda\mu}+\delta\cdot
H^{\alpha}_{\beta\lambda\mu}
\label{ddRiem}
\end{eqnarray}
where $H^{\alpha}_{\beta\lambda\mu}$ is a tensor called {\it the singular part
of the Riemann tensor distribution} and is defined only on the hypersurface as
follows
\begin{eqnarray}
H^{\alpha}_{\beta\lambda\mu} \equiv n_{\lambda} \left [ \Gamma^{\alpha}_{\beta
\mu} \right ] - n_{\mu} \left [ \Gamma^{\alpha}_{\beta\lambda} \right ] =
\frac{1}{2} \left\{ n^{\alpha} \left (
\zeta_{\beta\lambda}n_{\mu}-\zeta_{\beta\mu}n_{\lambda} \right)+n_{\beta}\left(
\zeta^{\alpha}_{\mu}n_{\lambda}-\zeta^{\alpha}_{\lambda}n_{\mu} \right )
\right\} \enspace . \label{Hache}
\end{eqnarray}
Of course, this tensor has the algebraic properties of a Riemann tensor. For
the Ricci tensor distribution, the analogous relation is
\begin{eqnarray}
{\underline{R}}_{\beta\mu}= \left( \1-\otheta\right)\cdot
R^{-}_{\beta\mu}+\otheta\cdot R^{+}_{\beta\mu} +\delta\cdot H_{\beta\mu}
\label{ddRicc}
\end{eqnarray}
where we have introduced its singular part
\begin{eqnarray}
H_{\beta\mu}\equiv H^{\alpha}_{\beta\alpha\mu}= \frac{1}{2} \left (
n^{\alpha}\zeta_{\alpha\beta}n_{\mu}+ n^{\alpha}\zeta_{\alpha\mu}n_{\beta}
-n^{\alpha}n_{\alpha}\zeta_{\beta\mu}-\zeta^{\alpha}_{\alpha}n_{\beta}n_{\mu}
\right ) \label{Hach}
\end{eqnarray}
which is a symmetric tensor defined only at points of $\Sigma$. With regard to
the scalar curvature distribution, we also have
\begin{eqnarray}
\underline{R}=\left (\1- \otheta \right )\cdot R^{+}+ \otheta \cdot R^{-}  +
\delta\cdot H
\label{ddec}
\end{eqnarray}
where its singular part, defined only on the hypersurface, is given by
\begin{eqnarray}
H \equiv \left. g^{\beta\mu} \right \vert_{\Sigma}H_{\beta\mu}=
n^{\alpha}n^{\beta}\zeta_{\alpha\beta}- n^{\alpha}n_{\alpha}
\zeta^{\beta}_{\beta} \enspace . \label{hac}
\end{eqnarray}
Finally, it is interesting to find a similar expression for the Einstein tensor
distribution, because of its close relation with the matter contents of the
spacetime. Using the previous formulas, the definition
$G_{\beta\mu}=R_{\beta\mu}-\frac{1}{2}g_{\beta\mu} R$ allows us to write
\begin{eqnarray}
{\underline{G}}_{\beta\mu}= \left( \1-\otheta
\right)\cdot G^{-}_{\beta\mu}+\otheta\cdot G^{+}_{\beta\mu}
+\delta\cdot\tau_{\beta\mu} \label{Einst}
\end{eqnarray}
where we have defined a new symmetric tensor on $\Sigma$ as
\begin{eqnarray}
\tau_{\beta\mu}\equiv H_{\beta\mu}-\frac{1}{2}\left .g_{\beta\mu}\right 
\vert_{\Sigma}H=\frac{1}{2}
\left \{ n^{\alpha}\zeta_{\alpha\beta}n_{\mu}+n^{\alpha}\zeta_{\alpha
\mu}n_{\beta}-n^{\alpha}n_{\alpha}\zeta_{\beta\mu}-\zeta^{\alpha}_{\alpha}
n_{\beta}n_{\mu}\right .\nonumber\\
\left .-\left .g_{\beta\mu}\right \vert_{\Sigma} \left(n^{\alpha}n^{\rho}\zeta_{\alpha\rho}
-n^{\alpha}n_{\alpha}\zeta^{\rho}_{\rho}\right ) \right \} .
\hspace{3cm} \label{eta}
\end{eqnarray}
Contracting the last expression with the normal vector to the hypersurface
$n^{\mu}$ we find that
\begin{eqnarray}
\tau_{\beta\mu}n^{\mu}=0 \enspace . \label{tausup}
\end{eqnarray}

Using the projection tensor $P^{\alpha}_{\beta}$, we can decompose the tensor
$\zeta_{\beta\mu}$, defined at every point of the hypersurface, into its
tangent part (with respect to the rigging vector $\vec{\r}\,$) and its rigged
part as
\begin{eqnarray}
\zeta_{\beta\mu}=\zeta^{\vec{\r}}_{\beta\mu}+\zeta^{\vec{\r}}_{\beta}n_{\mu}
+\zeta^{\vec{\r}}_{\mu}n_{\beta}+\zeta^{\vec{\r}}n_{\beta}n_{\mu}
\label{decomp}
\end{eqnarray}
where we have defined
\begin{eqnarray}
\zeta^{\vec{\r}}_{\beta\mu}\equiv P^{\lambda}_{\beta}P^{\alpha}_{\mu}
\zeta_{\lambda\alpha}\, , \hspace{0.3cm}
\zeta^{\vec{\r}}_{\beta}\equiv \r^{\lambda}P^{\rho}_{\beta}\zeta_{\lambda\rho}\, ,
\hspace{0.3cm} \zeta^{\vec{\r}}\equiv
\r^{\lambda}\r^{\mu}\zeta_{\lambda\mu}\enspace  . \label{defin}
\end{eqnarray}
Here, the first two objects are tangent to the rigged hypersurface in the sense
that they are orthogonal to the rigging vector. Therefore, they are
isomorphically related with tensors defined in the hypersurface through the
maps $\Phi^{\star}$ and $\Lambda$ of section 2.  The important point now is
that substituting the decomposition (\ref{decomp}) of $\zeta$ in expression
(\ref{Hache}) we find
\begin{eqnarray}
H_{\alpha\beta\lambda\mu}=\frac{1}{2}\left\{ n_{\alpha}\left (
\zeta^{\vec{\r}}_{\beta\lambda}n_{\mu}-\zeta^{\vec{\r}}_{\beta\mu}n_{\lambda}
\right)+n_{\beta}\left( {\zeta}^{\vec{\r}}_{\alpha\mu}n_{\lambda}-
{\zeta}^{\vec{\r}}_{\alpha\lambda}n_{\mu} \right ) \right\} \label{Ha}
\end{eqnarray}
so that only the tangent part $\zeta^{\vec{\r}}_{\beta\mu}$ of
$\zeta_{\beta\mu}$ appears in the singular part of the Riemann tensor
distribution.

Next, we shall find an intrinsic expression for $\zeta^{\vec{\r}}_{\beta\mu}$
depending only on the tensor ${\cal H}_{ab}$. To that end, let us recall that
its definition is ${\cal H}_{ab}=e_a^{\nu}e_b^{\mu}\nabla_{\nu}\r_{\mu}$ and,
consequently, being the connection discontinuous across $\Sigma$, ${\cal
H}_{ab}$ will be different when coming from $V^{+}$ or from $V^{-}$. Although
this object is defined only in the hypersurface and then it cannot be
continuous nor discontinuous, we will denote by $\left [{\cal H}_{ab} \right ]$
the difference at each point in $\Sigma$ of ${\cal H}_{ab}$ defined with the
connection of $V^{+}$ and ${\cal H}_{ab}$ defined with the connection of
$V^{-}$.\footnote{For other intrinsic objects of the hypersurface we will use
also the brackets to denote the difference between the objects defined form
$V^+$ and $V^-$. Abusing the language, we will name `continuous' such objects
with vanishing difference.} Making use of (\ref{Crissym}) this difference
tensor becomes
\begin{eqnarray}
\left [ {\cal H}_{ab} \right] =- \r_{\mu}\left [ \Gamma^{\mu}_{\sigma\nu}
\right ] e_a^{\nu}e_b^{\sigma}=- \frac{1}{2}\r_{\mu}\left(
\zeta^{\mu}_{\sigma}n_{\nu}+\zeta^{\mu}_{\nu}n_{\sigma}-n^{\mu}\zeta_{\sigma
\nu}
\right ) e_a^{\nu}e_b^{\sigma}=\frac{1}{2}\zeta_{\sigma\nu}e_a^{\nu}e_b^{\sigma}
\label{eich}
\end{eqnarray}
where we have taken into account that the rigging form $\r_{\mu}$, the tangent
vectors $e_a^{\mu}$ and $e_a^{\nu}\partial_{\nu}e^{\mu}_b$ have the same value
from $V^{-}$ or $V^{+}$. Note that $\left[ {\cal H}_{ab} \right ]$ is symmetric
despite the fact that ${\cal H}_{ab}$ itself is not. $\left[ {\cal H}_{ab}
\right ]$ is uniquely related, through the map $\Lambda$, with a symmetric
two-covariant tensor field in the manifold defined only at points on the
hypersurface. This tensor, which we will denote as before with the same symbol
$\left [{\cal H}_{\mu\nu}\right ]$, is
\begin{eqnarray*}
\left [{\cal H}_{\mu\nu}\right ]= 
\frac{1}{2}\zeta_{\sigma\rho}P^{\sigma}_{\mu}P^{\rho}_{\nu}=\frac{1}{2}
\zeta^{\vec{\r}}_{\mu\nu}
\end{eqnarray*}
and, of course, it satisfies
\begin{eqnarray}
\left [{\cal H}_{\mu\nu}\right ]\r^{\mu}=0. \label{norm}
\end{eqnarray}
Therefore, we obtain for the singular part of the Riemann tensor distribution
\begin{eqnarray}
H^{\alpha}_{\beta\lambda\mu}=n^{\alpha}\left ( -\left [{\cal H}_{\beta\mu}
\right ] n_{\lambda}+\left [ {\cal H}_{\beta\lambda} \right ] n_{\mu} \right )
+ n_{\beta} \left ( -\left [ {\cal H}^{\alpha}_{\lambda} \right ] n_{\mu} +
\left [{\cal H}^{ \alpha}_{\mu} \right ] n_{\lambda} \right ) \label{probth}
\end{eqnarray}
and this expression allows us to prove the following fundamental theorem
\begin{theorem}
The singular part of the Riemann tensor distribution vanishes if and only if
$\left [ {\cal H}_{\mu\nu} \right ]=0$, or equivalently, iff $\left [ {\cal
H}_{ab} \right]=0$.
\end{theorem}
{\it Proof:}
 
\noindent $\left [ \Rightarrow \right ]$ 
If $\left[ {\cal H}_{\alpha\beta} \right ]=0 $ then from the previous formula
$H^{\alpha}_{\beta\lambda\mu}=0$.

\noindent $\left [ \Leftarrow \right ]$ Suppose now that 
$H^{\alpha}_{\beta\lambda\mu}=0$. Contracting then (\ref{probth}) with
$\r^{\beta}\r^{\lambda}$ and making use of (\ref{norm}) we obtain $0=
\r^{\beta}\r^{\lambda}H^{\alpha}_{\beta\lambda\mu}=  \left [ {\cal
H}^{\alpha}_{\mu} \right ]$ and the theorem follows.

\vspace{0.5cm}

From now on, we shall refer to $\left [ {\cal H}_{\mu\nu} \right ]=0$ as {\it
the junction conditions}. These conditions assure that all the curvature (or
matter) tensors have, at most, finite discontinuities across $\Sigma$.
Furthermore, when the junction conditions are satisfied we get from
(\ref{der1}) and (\ref{decomp}) a structure for the discontinuities of the
first derivatives of the metric tensor which allows us to perform a $C^1$
change of coordinates such that the metric becomes $C^1$, in accordance with
the minimal differentiability requirements of Lichnerowicz
\cite{L}.
In order to see this, let us note that under $C^1$ change of coordinates,
$x^{\alpha'}=x^{\alpha'}\left( x^{\beta}\right )$ the discontinuity of the
partial derivative of the jacobian matrix of the transformation reads
\begin{eqnarray}
\left[\frac{\partial}{\partial x^{\mu'}}\left( \frac{\partial x^{\alpha}}{\partial
x^{\alpha'}}\right )\right ]=n_{\mu'}n_{\alpha'}B^{\alpha}
\end{eqnarray}
where $B^{\alpha}$ is a vector defined on the hypersurface.  Hence it is easy
to check from the transformation law for the discontinuities of the first
derivatives of the metric that if we choose a coordinate change satisfying
$B_{\alpha}=-\left(\zeta^{\vec{\r}}_{\alpha}+\frac{1}{2}\zeta^{\vec{\r}}
n_{\alpha}\right)$ then the metric becomes $C^1$ in the new coordinates.

Similar results can be proven for the singular parts of the Ricci tensor, the
scalar curvature and the Einstein tensor distributions. The expressions for
these singular parts come directly from (\ref{probth}) and read, respectively
\begin{eqnarray}
H_{\beta\mu}=-\left( \vec{n}\cdot\vec{n} \right ) \left [ {\cal H}_{\beta\mu}
\right ] + n^{\alpha}\left[ {\cal H}_{\alpha\beta} \right ] n_{\mu}
+ n^{\alpha}\left [{\cal H}_{\alpha\mu} \right ] n_{\beta} - \left [ {\cal
H}^{\alpha}_{\alpha} \right ] n_{\beta}n_{\mu}\enspace , \label{ricsing}
\end{eqnarray}
\begin{eqnarray}
H= -2\left(  \vec{n} \cdot \vec{n} \right ) \left [ {\cal H}_{\alpha}^{\alpha}
\right ] + 2 n^{\alpha}n^{\beta} \left [ {\cal H}_{\alpha\beta} \right ]
\enspace ,\label{Rsing} 
\end{eqnarray}
\begin{eqnarray} 
\tau_{\beta\mu}=-\left ( \vec{n} \cdot \vec{n} \right)  \left [ {\cal H}_{\beta
\mu} \right ]+ n^{\alpha} \left [ {\cal H}_{\alpha\beta} \right ] n_{\mu}
+n^{\alpha} \left [ {\cal H}_{\alpha\mu} \right ] n_{\beta} - \left [ {\cal
H}_{\alpha}^{\alpha} \right ] n_{\beta}n_{\mu} \nonumber \\ - \left
.g_{\beta\mu}\right \vert_{\Sigma}
\left ( -\left ( \vec{n} \cdot \vec{n} \right ) \left [ 
{\cal H}^{\alpha}_{\alpha}\right ]+ n^{\alpha}n^{\nu} \left [ {\cal
H}_{\alpha\nu}
\right ] \right ) \enspace .\hspace{2cm} \label{EMsing}
\end{eqnarray}
Summarizing, we have the following general theorem.
\begin{theorem}
\begin{enumerate}
\item At a point $x \in \Sigma$ where the hypersurface is not null, the singular 
part of the Ricci tensor distribution vanishes if and only if $\left [ {\cal
H}_{\beta\mu} \right ]=0$, hence, iff the singular part of the Riemann tensor
distribution vanishes.
\item At a point $x \in \Sigma$ where the hypersurface is null, the singular part
of the Ricci tensor distribution vanishes if and only if $n^{\alpha} \left [
{\cal H}_{\alpha\beta} \right ]=0$ and $ \left [{\cal H}^{\alpha}_{\alpha}
\right ]= 0$.
\item The singular part of the energy-momentum tensor distribution vanishes if and
only if so does the singular part of the Ricci tensor distribution.
\item The singular part of the curvature scalar distributon vanishes if and only 
if $\left( \vec{n} \cdot \vec{n} \right )
\left [ {\cal H}_{\alpha}^{\alpha} \right ] = \left [ 
{\cal H}_{\beta\mu} \right ]n^{\beta}n^{\mu}$.
\end{enumerate}
\end{theorem}
{\it Proof:} For the first two assertions, if $H_{\beta\mu}=0$ then we can
contract equation (\ref{ricsing}) with $\r^{\beta}$ and $\r^{\mu}$ to obtain
$\left [ {\cal H}^{\alpha}_{\alpha} \right ]=0$, and contracting now the same
equation only with $\r^{\beta} $ and using this last result we have $
n^{\alpha}\left[{\cal H}_{\alpha\mu}\right ]=0$. Thus, $H_{\beta\mu}= - \left (
\vec{n} \cdot \vec{n} \right ) \left [ {\cal H}_{\beta\mu} \right ] =0$ so that
the direct part of the theorem follows. The converse is trivial from equation
(\ref{ricsing}) itself.  The third assertion is then immediate from the
definition $\tau_{\beta\mu}= H_{\beta\mu}- \frac{1}{2} \left .g_{\beta\mu}
\right \vert_{\Sigma}H $.  Finally, the last part is a simple consequence of
equation (\ref{Rsing}).

\vspace{0.5cm}

We see from this theorem that, at points where $\Sigma$ is not null, the
vanishing of the matter singular part is equivalent to the vanishing of the
full Riemann singular part, whereas at points where $\Sigma$ is null this is
not the case.

The above theorems seem to depend on the rigging vector $\vec{\r}$, even though
we have not chosen this vector field on the hypersurface. Therefore, it would
be very interesting to see that the results do not depend, in fact, on the
rigging vector $\vec{\r}$. This can be established straightforwardly.
\begin{theorem}
The condition $\left [{\cal H}_{\alpha\beta}\right ]=0$ is invariant under
arbitrary changes of the rigging vector $\vec{\r}$.
\end{theorem}
{\it Proof:} Let us consider another rigging $\vec{\r'}$. Its difference tensor
$\left [{{\cal H}'}_{\alpha\beta}\right ]$ is, of course
\begin{eqnarray*}
\left [ {{\cal H}'}_{\alpha\beta} \right ] = \frac{1}{2} {P'}^{\mu}_{\alpha}
{P'}^{\nu}_{\beta} \zeta_{\mu\nu}
\end{eqnarray*}
where ${P'}^{\mu}_{\alpha}=\delta_{\alpha}^{\mu}-\r'^{\mu}n_{\alpha}$ is the
projection tensor associated to the rigging $\vec{\r'}$. Using now the
decomposition (\ref{decomp}) of $\zeta_{\mu\nu}$ and the fact that
${P'}^{\alpha}_{\mu}$ is a projector to the hypersurface in the sense that
${P'}^{\alpha}_{\mu}n_{\alpha}=0$, we obtain
\begin{eqnarray*}
\left [{{\cal H}'}_{\alpha\beta}\right ]=\frac{1}{2}{P'}^{\mu}_{\alpha}
{P'}^{\nu}_{\beta}\zeta^{\vec{\r}}_{\mu\nu}={P'}^{\mu}_{\alpha}
{P'}^{\nu}_{\beta} \left [ {\cal H}_{\mu\nu} \right ]
\end{eqnarray*}
This equation shows that if $\left [ {\cal H}_{\mu\nu} \right ]=0$ then $\left
[{{\cal H}'}_{\mu\nu} \right ] =0$, and vice versa due to the symmetric role
played by $\vec{\r}$ and $\vec{\r'}$ in this reasoning.

\vspace{0.5cm}

Let us study now what is the difference of the objects in the hypersurface,
like $\Psi^a_b$, $\varphi_a$, $\Gam^a_{bc}$, $\Gamma^a_{bc}$ or $K_{ab}$, when
calculated from $V^{+}$ or $V^{-}$. It is very easy from their definitions to
find the following results
\begin{eqnarray}
\left [ \Psi^a_b \right ] = \frac{1}{2}g^{ac}e_c^{\mu}e_b^{\nu}\zeta_{\mu\nu}
&=&g^{ac}e_c^{\mu}e_b^{\nu}\left [ {\cal H}_{\mu\nu} \right ]= g^{ac}\left [
{\cal H}_{cb} \right ] \enspace ,  \label{discontP}\\
\left [ \varphi_a \right ] = \frac{1}{2} n^b e_b^{\mu}e_a^{\nu}\zeta_{\mu\nu}
&=& n^{b} e_b^{\mu}e_a^{\nu} \left [ {\cal H}_{\mu\nu} \right ]= n^{b}\left [
{\cal H}_{ba} \right ] \enspace , \label{discontvar} \\
\left [ K_{ab} \right ]= \frac{1}{2} \left ( \vec{n}\cdot\vec{n} \right )
e_a^{\mu}e_b^{\nu} \zeta_{\mu\nu}&=&\left (\vec{n}\cdot\vec{n} \right )
e_a^{\mu}e_b^{\nu}\left [ {\cal H}_{\mu\nu} \right ]=
\left (\vec{n}\cdot\vec{n} \right )\left [ {\cal H}_{ab} \right ]
\enspace , \label{discontK} \\
\left [ \Gamma^a_{bc} \right ]=-
\frac{1}{2}n^{a}e_b^{\mu}e_c^{\nu}\zeta_{\mu\nu}&=& -n^a e_b^{\mu} e_c^{\nu}
\left[ {\cal H}_{\mu\nu} \right ]=
-n^a \left [ {\cal H}_{bc} \right ] \enspace ,  \label{discontGam} \\
\left [ \Gam^a_{bc} \right ]&=&0 \enspace . \label{discontGam2}
\end{eqnarray}
From (\ref{discontGam2}) we learn that the rigged metric connection is {\it
always}, and for any rigging, continuous. In this sense, the rigged metric
connection is more intrinsic for general hypersurfaces than the rigged
connection. Consequence of this, (or directly of equations (\ref{discontGam})
and (\ref{discontvar})), is that $\Gamma_c$ has always the same definitions
from both sides of the hypersurface, as is trivial from its expression
$\Gamma_c=\partial_c \log \left( \eta_{123} \right )$.  The rigged connection,
as we can see from (\ref{discontGam}), is discontinuous in general unless
either we have already matched properly such that the junction conditions hold
or in the case of non-null hypersurfaces when the rigging is chosen canonically
as the normal vector (in which case we have $n^a=0$).

With regard to the second fundamental form, from (\ref{discontK}) we have the
standard and very well-known result (see, for instance, \cite{I},\cite{CD},
\cite{BI}):
\begin{theorem}
At any point $x \in \Sigma$ where the hypersurface is non-null, the necessary
and sufficient condition for the singular part of the Riemann tensor to vanish
is that the second fundamental form be continuous across the hypersurface.

On the other hand, the second fundamental form is always continuous at any
point $x \in \Sigma $ where the hypersurface is null.
\end{theorem}
This theorem has its counterpart in the case of general hypersurfaces, but we
must use the tensor $\Psi^a_b$ instead of the second fundamental form, and also
we have to choose a non-null rigging. The precise statement, which follows
directly from equation (\ref{discontP}) and proposition 4, is
\begin{theorem}
At any point $x \in \Sigma$, the necessary and sufficient condition for the
singular part of the Riemann tensor to vanish is that the tensor $\Psi^a_b$,
constructed with any non-null rigging, be the same defined from the region
$V^{+}$ and from the region $V^{-}$.
\end{theorem}

\vspace{0.3cm}

Next, we are going to show that the second Bianchi identity holds in the
distributional sense independently of whether or not the junction conditions
are satisfied (of course, the preliminary junction conditions and a $C^3$
differentiability for $g^{\pm}$ are both assumed).  To prove it, we do not have
to extend any object in the hypersurface outside from it, but we need to know
the covariant derivative of the Riemann tensor distribution. Although in the
general case we cannot define the covariant derivative of arbitrary
distributions because the connection symbols are discontinuous across the
hypersurface, for the Riemann distribution this can be done without problems at
once. This is due to the fact that the Riemann distribution components are able
to act on the following set of test functions: those with a well-defined
restriction to the hypersurface and such that they are {\it just} locally
integrable both in the spacetime and in the hypersurface. From definition
(\ref{covdis}) we can find after some simple calculations
\begin{eqnarray}
\left < \nabla_{\mu} {\underline{R}}^{\alpha}_{\beta\gamma\delta}, Y \right >
= \int_{V^{+}} \nabla_{\mu}{R^{+}}^{\alpha}_{\beta\gamma\delta} Y
\mbox{\boldmath$\eta$} + \int_{V^{-}} \nabla_{\mu} 
{R^{-}}^{\alpha}_{\beta\gamma\delta} Y \mbox{\boldmath$\eta$} + \int_{\Sigma}
n_{\mu}\left [ R^{\alpha}_{\beta\gamma\delta} \right ]Y d\sigma + \nonumber \\
+\int_{\Sigma} \left\{ \left ( -
H^{\alpha}_{\sigma\gamma\delta}\Gamma^{\sigma}_{\beta\mu}- H^{\alpha}_{\beta
\sigma\delta}\Gamma^{\sigma}_{\gamma\mu}-H^{\alpha}_{\beta\gamma\sigma}
\Gamma^{\sigma}_{\delta\mu} + H^{\nu}_{\beta\gamma\delta} \Gamma^{\alpha}_{\nu
\mu}-H^{\alpha}_{\beta\gamma\delta} \Gamma^{\rho}_{\mu\rho} \right ) Y 
- H^{\alpha}_{\beta\gamma\delta} \partial_{\mu} Y\right\} d\sigma
\, .\label{Bian1}
\end{eqnarray} 
From here it follows that we must know the discontinuity of the Riemann tensor
across the hypersurface. Equation (\ref{partderf}) applied to the Christoffel
symbols gives
\begin{eqnarray*}
\left [ \partial_{\gamma} \Gamma^{\alpha}_{\beta\delta} \right ]=
n_{\gamma} A^{\alpha}_{\beta\delta}+ \omega^a_\gamma\partial_a
\left [ \Gamma^{\alpha}_{\beta\delta} \right ] \enspace ,
\end{eqnarray*}
while the discontinuity $ \left [ \Gamma^{\alpha}_{\gamma
\rho}\Gamma^{\rho}_{\beta\delta} \right ] $ is easily found to be
\begin{eqnarray*}
\left [ \Gamma^{\alpha}_{\gamma\rho} \Gamma^{\rho}_{\beta\delta} \right ]=
\left.\Gamma^{\alpha}_{\gamma\rho}\right \vert_{\Sigma} 
\left [ \Gamma^{\rho}_{\beta\delta} \right ] + 
\left [ \Gamma^{\alpha}_{\gamma\rho} \right ] 
\left.\Gamma^{\rho}_{\beta\delta}\right \vert_{\Sigma} \enspace .
\end{eqnarray*}
Thus, we explicitly have (\ref{Bian1}) in terms of known objects and we can
evaluate
\begin{eqnarray}
\nabla_{\mu}{\underline{R}}^{\alpha}_{\beta\gamma\delta}+\nabla_{\gamma}
{\underline{R}}^{\alpha}_{\beta\delta\mu}+\nabla_{\delta}
{\underline{R}}^{\alpha}_{\beta\mu\gamma} \enspace .\label{Bian2}
\end{eqnarray}
After a long but straightforward calculation involving a decomposition of the
partial derivative of the test function $\partial_{\mu}Y = n_{\mu}
\r^{\nu}\partial_{\nu} Y + \omega_{\mu}^a \partial_a Y$, a careful integration
by parts and using the fact that
\begin{eqnarray*}
n_{\mu} H^{\alpha}_{\beta\gamma\delta}+n_{\gamma}H^{\alpha}_{\beta\delta\mu}
+n_{\delta}H^{\alpha}_{\beta\mu\gamma}=0
\end{eqnarray*}
as follows from the explicit expression (\ref{Hache}) of
$H^{\alpha}_{\beta\gamma\delta}$, together with the Bianchi identities for the
Riemann tensors in $V^{+}$ and $V^{-}$, we find that expression (\ref{Bian2})
applied to a test function Y produces the following result
\begin{eqnarray*}
\int_{\Sigma} \left \{ \left [ \Gamma^{\alpha}_{\beta\delta} \right ]
\left (  \omega^a_{\mu}\partial_a n_{\gamma} - \omega^a_{\gamma} \partial_a
n_{\mu} + n_{\gamma} ( \partial_a \omega^a_{\mu} + \omega^a_{\mu}
\Gamma_a ) - n_{\mu} ( \partial_a \omega^a_{\gamma} + \omega^a_{\gamma}
\Gamma_a )\right ) - \Gamma^{\nu}_{\mu\nu} H^{\alpha}_{\beta\gamma
\delta} + {\rm c.t.} \right \}d\sigma
\end{eqnarray*}
where c.t. represents the two other terms obtained from the one shown by
permuting ciclically the indexes $\mu$, $\gamma$ and $\delta$.  Taking into
account the formulas of previous sections, it is not difficult to see that
\begin{eqnarray*}
\omega^a_{\mu}\partial_a n_{\gamma} - \omega^a_{\gamma} \partial_a
n_{\mu} + n_{\gamma} ( \partial_a \omega^a_{\mu} + \omega^a_{\mu}
\Gamma_a ) - n_{\mu} ( \partial_a \omega^a_{\gamma} + \omega^a_{\gamma}
\Gamma_a )=n_{\gamma} \Gamma^{\nu}_{\mu\nu} - n_{\mu} \Gamma^{\nu}_{\gamma\nu}
\end{eqnarray*}
and recalling again expression (\ref{Hache}) we arrive then to the Bianchi
identities
\begin{eqnarray}
\nabla_{\mu}{\underline{R}}^{\alpha}_{\beta\gamma\delta}+\nabla_{\gamma}
{\underline{R}}^{\alpha}_{\beta\delta\mu}+\nabla_{\delta}
{\underline{R}}^{\alpha}_{\beta\mu\gamma}=0 \label{Bianchi}.
\end{eqnarray}
Of course, from this we also have
\begin{eqnarray*}
\nabla_{\mu}\underline{G}^{\mu\nu}=0 \enspace ,
\end{eqnarray*}
so that, when the metric is continuous, the energy-momentum tensor distribution
is conserved in the distributional sense.
\section{Physical Implications of the Junction Conditions.}
Let us now assume that we have made a proper matching between spacetimes such
that the singular part of the Riemann tensor distribution vanishes or,
equivalently, for every rigging vector field $\vec{\r}$ we have $\left [ {\cal
H}_{\mu\nu} \right ]=0$.  Our aim is to find the allowable discontinuities of
the Riemann tensor, that is to say, we want to know which physical components
of the curvature, i.e. of the Einstein tensor (matter contents) and the Weyl
tensor (pure gravitational field), can have discontinuities across the
hypersurface. Given that $\left [ {\cal H}_{\mu\nu} \right ]=0$, we get from
formulas (\ref{discontP}-\ref{discontGam}) above that
\begin{eqnarray*}
\left [ \Psi^a_b \right ]=0 \, , \hspace{0.2cm} \left [ \varphi_c \right ]=0 
\, , \hspace{0.2cm} \left [ K_{ab} \right ]=0 \, , \hspace{0.2cm}
\left [ \Gamma^a_{bc} \right ]=0 
\end{eqnarray*}
and therefore the covariant derivative $\nab$ in the hypersurface has the same
definition coming from $V^{+}$ or $V^{-}$. Using now the Gauss equation
(\ref{Gauss}) and the three Codazzi equations
(\ref{Cod-1},\ref{Cod-2},\ref{Cod-3}) we immediately find
\begin{eqnarray*}
\left [ \omega^d_{\alpha}R^{\alpha}_{\beta\gamma\delta}e_a^{\beta}e_b^{\gamma}
e_c^{\delta} \right ] =0 \, , &
\left [ n_{\mu}R^{\mu}_{\alpha\beta\gamma} 
e_a^{\alpha}e_b^{\beta}e_c^{\gamma} \right ] =0 \, , \\
\left [ \omega^c_{\mu}R^{\mu}_{\alpha\beta\gamma}\r^{\alpha}e_ a^{\beta}
e_b^{\gamma} \right ]=0 \, , &
\left [ n_{\mu}R^{\mu}_{\alpha\beta\gamma}\r^{\alpha}
e_a^{\beta}e_b^{\gamma}\right ]=0 \, .
\end{eqnarray*}
Noting that $\left \{ \n, {\mbox{\boldmath $\omega$}}^c \right \}$ and $\left
\{\vec{\r},{\vec{e}}_a\right \}$ are bases of their respective tangent spaces,
we have
\begin{theorem}
If the junction conditions $\left [ {\cal H}_{\mu\nu} \right ]=0$ are
satisfied, then
\begin{eqnarray}
\left [ R^{\mu}_{\alpha\beta\gamma}\right ] e_a^{\beta}e_b^{\gamma} =0 \enspace .
\label{necgen}
\end{eqnarray}
These are fourteen independent relations so that fourteen out of the twenty
components of the Riemann tensor must be continuous. Thus, only six independent
discontinuities in the curvature are allowed.
\end{theorem}
We can rewrite equations (\ref{necgen}) in several different forms. First of
all, given three independent vectors ${\vec{e}}_a$ and their normal one-form
$\n$, it is a general result for an object $(V)$ with two covariant
antisymmetric indexes that
\begin{eqnarray*}
{\left (V\right )}_{\lambda\mu}e_a^{\lambda}e_b^{\mu}=0 \Leftrightarrow
n_{\sigma} {\left (V \right )}_{\lambda\mu}+n_{\lambda} {\left (V
\right )}_{\mu\sigma}+n_{\mu} {\left (V \right )}_{\sigma\lambda}=0
\end{eqnarray*}
where we have written $(V)$ because this object can have more than two indexes.
Therefore the continuity conditions for the Riemann tensor can be rewritten as
\begin{eqnarray}
n_{\sigma}\left [R^{\alpha}_{\beta\lambda\mu}\right ]+n_{\lambda}\left [
R^{\alpha}_{\beta\mu\sigma}\right ]+ n_{\mu}\left
[R^{\alpha}_{\beta\sigma\lambda}\right]=0 \enspace . \label{cont1}
\end{eqnarray}
It is a known result that (\ref{cont1}) is still equivalent to the existence of
a  symmetric two-covariant tensor $B_{\mu\nu}$ such that
\begin{eqnarray}
\left [ R_{\alpha\beta\lambda\mu} \right ]=n_{\alpha}n_{\lambda}B_{\beta\mu}
-n_{\lambda}n_{\beta}B_{\alpha\mu}-n_{\mu}n_{\alpha}B_{\beta\lambda}+
n_{\mu}n_{\beta}B_{\alpha\lambda} \label{Bes},
\end{eqnarray}
where $B_{\lambda\mu}$ is a tensor field defined at the hypersurface and unique
up to a transformation of the type
\begin{eqnarray*}
{B'}_{\lambda\mu}=B_{\lambda\mu}+X_{\lambda}n_{\mu}+n_{\lambda}X_{\mu}
\end{eqnarray*}
with $X_{\lambda}$ an arbitrary one-form. The ten independent components of
$B_{\lambda\mu}$ minus the four gauge freedoms $X_{\lambda}$ give the six
arbitrary possible discontinuities of the Riemann tensor.  Direct consequences
of (\ref{cont1}) are the following
\begin{eqnarray}
n_{\sigma}\left [ R^{\sigma}_{\beta\lambda\mu} \right ]= n_{\lambda} \left [
R_{\beta\mu} \right ] - n_{\mu} \left [ R_{\beta\lambda} \right ] \enspace ,
\label{cont2} \\
n_{\sigma}\left [ R^{\sigma}_{\beta\lambda\mu} \right ]e_a^{\lambda}=-
n_{\mu}\left [R_{\beta\lambda} \right ] e_a^{\lambda} \label{cont3}
\end{eqnarray}
where we have not lost any information. In other words, equations
(\ref{necgen}), (\ref{cont1}), (\ref{Bes}), (\ref{cont2}) and (\ref{cont3})
are, all of them, equivalent to each other.

We are now ready to study which of the six allowable curvature discontinuities
are matter discontinuities or pure gravitational (Weyl) ones.  Contracting
indexes ${\beta}$ and ${\mu}$ in equation (\ref{cont2}) we get
\begin{eqnarray}
n^{\sigma}\left [R_{\sigma\lambda}\right ]=\frac{1}{2}n_{\lambda}\left [R\right
] \label{isr}
\end{eqnarray}
or, what is the same,
\begin{eqnarray}
n^{\sigma}\left [G_{\sigma\lambda} \right ]= 0 \label{eins}
\end{eqnarray}
where, of course, $G_{\sigma\lambda}$ is the Einstein tensor of the manifold.
These relations were known in the case of non-null junction hypersurfaces as
the Israel conditions \cite{I}. This equation tells us that four components of
the Einstein tensor cannot have any discontinuities across the junction
hypersurface. In order to see which are the remaining ten continuous components
of the curvature, we use the decomposition of the Riemann tensor in terms of
the Weyl tensor, Ricci tensor and scalar curvature
\begin{eqnarray}
R^{\alpha}_{\beta\lambda\mu}=C^{\alpha}_{\beta\lambda\mu}+\frac{1}{2} \left(
R^{\alpha}_{\lambda}g_{\beta\mu}-R^{\alpha}_{\mu}g_{\beta\lambda}+
\delta^{\alpha}
_{\lambda}R_{\beta\mu}-\delta^{\alpha}_{\mu}R_{\beta\lambda} \right ) -
\frac{1}{6} R \left (\delta^{\alpha}_{\lambda}g_{\beta\mu}-\delta^{\alpha}_{\mu}
g_{\beta\lambda}
\right ) \label{RWri}
\end{eqnarray}
and then we can rewrite equations (\ref{cont2}) and (\ref{cont3}),
respectively, in the the following form
\begin{eqnarray}
n_{\sigma}\left [ C^{\sigma}_{\beta\lambda\mu}\right ]=\frac{1}{12}
\left [ R \right ] \left ( g_{\beta\lambda}n_{\mu}- g_{\beta\mu}n_{\lambda}
\right)
+\frac{1}{2} \left ( n_{\lambda} \left [ R_{\beta\mu} \right ] - n_{\mu}
\left [ R_{\beta\lambda} \right ] \right ) \enspace , \label{weyl1} \\
n_{\sigma} \left [C^{\sigma}_{\beta\lambda\mu}\right ]e_a^{\lambda}=
\frac{1}{12}
n_{\mu} e_a^{\lambda}\left ( \left [ R \right ]g_{\beta\lambda}-6\left [
R_{\beta\lambda} \right ] \right ) \hspace{16mm} \label{weyl2}
\end{eqnarray}
where we have used (\ref{isr}). Despite of this, these two relations are again
fully equivalent to the general condition (\ref{necgen}).  From the last
equation it is immediate that
\begin{eqnarray}
n_{\sigma} \left [C^{\sigma}_{\beta\lambda\mu} \right ] e_a^{\lambda} e_b^{\mu}
=0 \label{weylp}
\end{eqnarray}
and this equation contains five independent relations so that these five out of
the ten components of the Weyl tensor must be continuous across the
hypersurface.  We have hitherto decomposed the fourteen conditions on the
Riemann tensor into four on the Einstein tensor (or Ricci tensor) and five on
the Weyl tensor. We will complete these relations by identifying the possible
discontinuities of the Riemann tensor and writing down all the components of
the Ricci and Weyl tensors in terms of these discontinuities. From formula
(\ref{Bes}) it is clear that the independent discontinuities of the Riemann
tensor are
\begin{eqnarray}
\r^{\alpha}e_a^{\beta}\r^{\lambda}e_b^{\mu}\left [R_{\alpha\beta\lambda\mu}
\right]= e_a^{\beta}e_b^{\mu}B_{\beta\mu}\equiv B_{ab}    ,\label{discon-}
\end{eqnarray} 
where obviously $B_{ab}$ are independent of the gauge freedom of $B_{\alpha
\beta}$ and are the six independent allowed discontinuities in the Riemann tensor.
The discontinuities of the Ricci and Weyl tensor can be straightforwardly
written in terms of these quantities using equation (\ref{Bes}) and the
decomposition (\ref{RWri}) as
\begin{eqnarray}
\left [R\right ] &=& 2\left (\vec{n}\cdot \vec{n}\right )g^{ab}B_{ab}
-2n^{a}n^{b}B_{ab} \label{compl1} \\
\r^{\alpha}\r^{\beta}\left [R_{\alpha\beta}\right ]&=&g^{ab}B_{ab}
\label{compl2} \\
\r^{\alpha}e_b^{\beta}\left[R_{\alpha\beta}\right]&=&-n^aB_{ab}
\label{compl3} \\
e_a^{\alpha}e_b^{\beta}\left[R_{\alpha\beta}\right]&=& \left (\vec{n}\cdot
\vec{n}\right )
B_{ab} \label{compl4} \\
\r^{\alpha}e_a^{\beta}\r^{\lambda}e_b^{\mu} \left [C_{\alpha\beta\lambda\mu}
\right] &=& \frac{2-\left(\vec{n}\cdot\vec{n}\right)\left(\vec{\r}\cdot
\vec{\r}\,\right)}{2}B_{ab}-\frac{1}{2}n^c\left(
B_{cb}\r_a+B_{ca}\r_b\right)-\frac{\r_a\r_b}{3}\left(\left ( \vec{n}\cdot
\vec{n}\right )g^{cd}B_{cd}\right. \nonumber \\
& &\left. -n^cn^dB_{cd}\right) -\g_{ab}\left \{\frac{\vec{\r}\cdot\vec{\r}}{3}
n^cn^dB_{cd} + \frac{3-2\left(\vec{n}\cdot\vec{n}\right)\left(\vec{\r}
\cdot\vec{\r}\,\right)}{6}g^{cd}B_{cd}\right\} \label{compl5} \\
\r^{\alpha}e_b^{\beta}e_c^{\lambda}e_d^{\mu} \left [ C_{\alpha\beta\lambda\mu}
\right ]&=& \frac{1}{2}n^{a}\left(B_{ac}\g_{bd}-B_{ad}\g_{bc}\right)+
\frac{\vec{n}\cdot\vec{n}}{2}\left(B_{bc}\r_d-B_{bd}\r_{c}\right) \nonumber \\
& & +\frac{1}{3}
\left(\left (\vec{n}\cdot\vec{n}\right )g^{ef}B_{ef}-n^{e}n^{f}B_{ef}\right)
\left(\g_{bd}
\r_c -\g_{bc}\r_d\right) \label{compl6} \\
e_a^{\alpha}e_b^{\beta}e_c^{\lambda}e_d^{\mu}\left[C_{\alpha\beta\lambda\mu}
\right]&=& \frac{\vec{n}\cdot\vec{n}}{2}\left(B_{ad}\g_{bc}-B_{ac}\g_{bd}+
B_{bc}\g_{ad}-B_{bd}\g_{ac}\right) \nonumber \\ & & +\frac{1}{3}\left(\left
(\vec{n}\cdot\vec{n}\right )
g^{ef}B_{ef}-n^{e}n^{f}B_{ef}\right)\left(\g_{ac}\g_{bd}-\g_{ad}\g_{bc}\right)
\label{compl7}.
\end{eqnarray}
These equations contain both the Israel conditions (\ref{eins}) and the
relations concerning only the Weyl tensor (\ref{weylp}) and include all the the
information about the continuities of the Riemann tensor.  The question now is
to rewrite these equations in terms of six components of the Weyl or Ricci
tensor which can be arbitrarily discontinous across the hypersurface. For an
everywhere non-degenerate hypersurface these components can be chosen as the
tangent part of the Ricci or Einstein tensor, as is clear from relation
(\ref{compl4}), but for a general hypersurface those are not arbitrary because
they must tend to zero in the singular points where the hypersurface is
degenerate.  A suitable set of six independent components whose discontinuities
are arbitrary everywhere in a general hypersurface is
\begin{eqnarray}
\r^{\alpha}\r^{\beta}\left [R_{\alpha\beta}\right] - 
\frac{\vec{\r}\cdot\vec{\r}}{3}\left [ R \right ] \equiv \Omega
 \hspace{1cm}, \hspace{1cm}
\r^{\alpha}e_a^{\beta}\r^{\lambda}e_b^{\mu}\left [ C_{\alpha\beta\lambda\mu} 
\right ]\equiv S_{ab}. \label{indep}
\end{eqnarray}
The six components $\r^{\alpha}e_b^{\beta}\r^{\lambda}e_c^{\mu}\left [
C_{\alpha\beta\lambda\mu}\right ]$ satisfy the relation
\begin{eqnarray*}
g^{ab}S_{ab} \equiv g^{ab}\r^{\alpha}e_b^{\beta}\r^{\lambda}e_c^{\mu}\left [
C_{\alpha\beta\lambda\mu}\right ]=0 ,
\end{eqnarray*}
consequence of the fact that the Weyl tensor is traceless. So the components
written above constitute really a set of six independent quantities.

From equation (\ref{compl5}) it is clear that $B_{ab}$ can be written in terms
of these components when $\left (\vec{n}\cdot
\vec{n}\right)\left(\vec{\r}\cdot\vec{\r}\,\right ) \neq 2$.
\begin{eqnarray}
B_{ab}=\frac{2}{2-\left(\vec{n}\cdot\vec{n}\right)\left(\vec{\r}\cdot\vec{\r}
\,\right)}\left( S_{ab} + n^c S_{cb} \r_a + n^c S_{ca} \r_b + \frac{1}{2}
\g_{ab}\Omega \right ) . \label{aislB}
\end{eqnarray}
This is not a restriction because this is a condition only on the rigging but
not on the point type of the hypersurface. We can always choose a rigging
vector $\vec{\r}$ that satisfies the above inequality. Thus, in terms of the
quantities (\ref{indep}) the fourteen continuity relations are
\begin{eqnarray}
\left [ R \right ] &=& \frac{6}{2-\left( \vec{n}\cdot\vec{n} \right )\left(
 \vec{\r}\cdot\vec{\r}\, \right )}\left ( -2n^en^d S_{ed}
+\left(\vec{n}\cdot\vec{n} \right) \Omega \right )\label{fina1} \\
\r^{\alpha}\r^{\beta}\left[ R_{\alpha\beta} \right ] &=& \frac{
\left(2+\left( \vec{n}\cdot\vec{n}\right )\left(\vec{\r}\cdot
\vec{\r}\,\right )\right)\Omega -4 \left(\vec{\r}\cdot\vec{\r}\,\right) n^en^d
S_{ed}}
{2-\left(\vec{n}\cdot\vec{n}\right)\left(\vec{\r}\cdot\vec{\r}\,\right)}
\label{fina2} \\
\r^{\alpha}e_c^{\beta}\left [R_{\alpha\beta} \right ] &=& -2 n^b S_{bc}
+ \frac{\r_c}{2-\left( \vec{n}\cdot\vec{n} \right )\left( \vec{\r}
\cdot\vec{\r}\, \right )}\left (  
\left(\vec{n}\cdot\vec{n}\right ) \Omega -2n^en^dS_{ed} \right )
\label{fina3} \\
e_a^{\alpha}e_b^{\beta}\left [R_{\alpha\beta}\right ] &=& \frac{\vec{n}
\cdot\vec{n}}{2-\left( \vec{n}\cdot\vec{n} \right )\left( \vec{\r}
\cdot\vec{\r}\, \right )} \left( 2S_{ab} +2\left( n^e S_{eb}
\r_a + n^e S_{ea} \r_b \right )+ \g_{ab} \Omega \right ) \label{fina4} \\
\r^{\alpha}e_a^{\beta}e_b^{\lambda}e_c^{\mu}\left [ C_{\alpha\beta\lambda\mu}
\right ] &=& \left ( \g_{ac}n^{d} S_{bd}- \g_{ab} n^d S_{dc} \right )
- \frac{ n^dn^e S_{de} \left(\g_{ac}\r_b-\g_{ab}\r_c\right )}
{2-\left(\vec{n}\cdot\vec{n}\right )\left(\vec{\r}\cdot\vec{\r}\,
\right)} \nonumber \\
& &+\frac{\left( \vec{n}\cdot\vec{n}\right )
\left \{ \r_a\left (\r_c n^d S_{db}- \r_b n^d S_{dc}\right )-
\left( \r_b S_{ac} - \r_c S_{ab} \right ) \right \}}
{2-\left(\vec{n}\cdot\vec{n}\right )\left(\vec{\r}\cdot\vec{\r}\,
\right)} \label{fina5}\\
e_a^{\alpha}e_b^{\beta}e_c^{\lambda}e_d^{\mu}\left [C_{\alpha\beta\lambda\mu}
\right ] &=& \frac{\vec{n}\cdot\vec{n}}{2-\left(\vec{n}\cdot\vec{n}\right)
\left(\vec{\r}\cdot\vec{\r}\,\right)}
\left \{ \g_{bc} S_{ad}- \g_{bd} S_{ac}+ \g_{ad} S_{bc}- \g_{ac} S_{bd}
\right .\nonumber \\ 
& &\left.+ n^e S_{ea}\left (\r_d\g_{bc} - \r_c\g_{bd} \right )  + n^e S_{eb}
\left ( \r_c\g_{ad} - \r_d\g_{ac} \right ) \right.  \nonumber \\
& & \left.+ n^e S_{ed}\left (\r_a\g_{bc}-\r_b\g_{ac}\right ) + n^e S_{ec}
\left (\r_b\g_{ad}-\r_a\g_{bc} \right ) \right \} \nonumber \\
& & + \frac{2n^en^fS_{ef}}{2-\left(\vec{n}\cdot\vec{n}\right)
\left(\vec{\r}\cdot\vec{\r}\,\right)}
\left (\g_{ad}\g_{bc}-\g_{ac}\g_{bd} \right ) .
\label{fina6}
\end{eqnarray}
Several interesting considerations can be made at the sight of these equations
about the behaviour of the discontinuities of the Ricci and Weyl tensors across
the junction hypersurface of two properly matched spacetimes, but as they are
self-evident from the equations we will not discuss them unless for the most
important ones. We write them down in the form of
\begin{theorem}

If the junction conditions are verified on the matching hypersurface, the
following properties about the discontinuities of the Riemann tensor hold:

\begin{enumerate}

\item The four normal components (\ref{eins}) of the energy-momentum tensor
 are necessarily continuous across the hypersurface (Israel conditions).
 
\item The five components (\ref{weylp}) of the Weyl tensor are continuous
across the matching hypersurface.
 
\item A suitable set of six independent allowable discontinuities of the Ricci
 and Weyl tensors for general hypersurfaces which change in character is given
by (\ref{indep}).
 
\item At any point $x \ \in \ \Sigma$  where the matching hypersurface is
  non-null the six independent allowable discontinuities can be chosen as
arbitrary matter discontinuities given by $\left
[T_{\mu\nu}e_a^{\mu}e_b^{\nu}\right ]$.

\item For null points the tangent part of the Ricci tensor must be continuous
  (see (\ref{compl4})) across the hypersurface. Moreover, four of the six
independent allowable discontinuities can be chosen as matter discontinuities
given by $\left[\r^{\mu}R_{\mu\nu}\right]$ and the other two arbitrary
discontinuities are given by the spinorial component $[\Psi_4]$ in any null
tetrad with $\vec{n}$ as the first real null vector.
  
\item Near the singular points of the hypersurface the tangent components of the
    Ricci tensor must tend to zero at least as the first power of the norm of
the normal vector (see (\ref{compl4})).
   
\end{enumerate}
\end{theorem}

To end this section we consider the question if there exists any differential
equation governing the evolution of the arbitrary discontinuities of the
Riemann tensor. The Bianchi identities, which are true both in $V^{+}$ and $
V^{-}$ imply obviously that
\begin{eqnarray}
\left [ \nabla_{\rho}R_{\alpha\beta\lambda\mu} + \nabla_{\lambda}R_{\alpha
\beta\mu\rho}+ \nabla_{\mu}R_{\alpha\beta\rho\lambda} \right ] =0.\label{BB}
\end{eqnarray}
Due to the general formula (\ref{partderf}), the discontinuity of the covariant
derivative of the Riemann tensor can be written
\begin{eqnarray*}
\left [ \nabla_{\rho}R_{\alpha\beta\lambda\mu} \right ] = n_{\rho}
t_{\alpha\beta\lambda\mu} + \omega_{\rho}^a \left ( e_a^{\sigma}\nabla_{\sigma}
\left [R_{\alpha\beta\lambda\mu}\right ]\right ) ,
\end{eqnarray*}
where $t_{\alpha\beta\lambda\mu}$ is an arbitrary tensor on the hypersurface
with the symmetries of a Riemann tensor. Substituting in the second term of
this relation the equation for the discontinuity of the Riemann tensor
(\ref{Bes}) and using some of the relations written down in a previous section
it can be proven that the Bianchi discontinuity relation (\ref{BB}) is
completely equivalent to the relation
\begin{eqnarray}
t_{\alpha\beta\lambda\mu}e_b^{\lambda}e_d^{\mu}&=& \frac{1}{2}
\left ( K_{ad}B_{bc}-K_{cd}B_{ba}
-K_{ab}B_{dc}+K_{cb}B_{da}
\right )\left ( \omega^a_{\alpha}\omega^c_{\beta}-\omega^a_{\beta}
\omega^c_{\alpha}\right ) \nonumber \\
& &+ \left ( \nab_bB_{ad}-\nab_dB_{ab}-2\varphi_bB_{ad}+2\varphi_dB_{ab}\right)
\left( n_{\alpha}\omega^a_{\beta}-n_{\beta}\omega^a_{\alpha} \right).
\label{derriem}
\end{eqnarray}
As before, these are fourteen relations for the twenty independent components
of $t_{\alpha\beta\lambda\mu}$, so there appear six new arbitrary independent
discontinuities in $\nabla_{\rho}R_{\alpha\beta\lambda\mu}$.  The other six
relations contained in the Bianchi identities are identically verified for the
derivatives of the discontinuity of the Riemann tensor and do not involve the
tensor $t_{\alpha\beta\lambda
\mu}$. Thus the
Bianchi discontinuity relations do not provide us with any evolution equation
for the discontinuity of the Riemann tensor. However, we see from
(\ref{derriem}) that the discontinuities in the first derivatives of the
Riemann tensor (apart from the six new arbitrary ones) involve not only the
discontinuities $B_{ab}$ of the Riemann tensor itself and its first
derivatives, but also intrinsic properties of the matching hypersurface,
namely, the second fundamental form and the one-form {\boldmath$\varphi$}.

\section{Acknowledgements}
The present work has been partially supported by DGICYT project
PB90-0482-C02-01. M.M. wishes to thank the Direcci¢ General d'Universitats,
Generalitat de Catalunya for financial support.

\section{Appendix.}
In order to define tensor distributions (see, for instance \cite{T},\cite{Li})
it is necessary to construct the set ${\cal D}(V_4)$ of test tensor fields,
that is, the set of $C^{\infty}$ tensor fields of any order with compact
support in the manifold. We denote by ${\cal D}^q_p$ the subset of p-covariant
q-contravariant tensor fields with compact support. Then, p-covariant
q-contravariant tensor distributions $\chi^q_p$ are defined as linear and
continuous functionals from ${\cal D}_q^p$ to the real numbers, that is to say
\begin{eqnarray*}
\chi^q_p : {\cal D}^p_q & \rightarrow & {\rm I\!R} \\
Y^p_q & \rightarrow & \chi^q_p \left ( Y^p_q \right ) \equiv \left < \chi^q_p,
Y^p_q \right > \enspace .
\end{eqnarray*}
The sum of two tensor distributions of the same type and the product of a
tensor distribution by a real number can be defined in the usual way. With
these definitions the space of tensor distributions of a given type constitutes
a vector space. Given any locally integrable p-covariant q-contravariant tensor
field $T^q_p$ in an oriented manifold, we can define a p-covariant
q-contravariant tensor distribution associated to it as follows
\begin{eqnarray*}
{\underline{T}}^q_p : {\cal D}_q^p & \rightarrow & {\rm I\!R} \\ Y^p_q  &
\rightarrow & \left <{\underline{T}}^q_p, Y^p_q \right >\equiv
\int_{V_4} \left ( T,Y \right ) \mbox{\boldmath$\eta$}
\end{eqnarray*}
where $\left (T,Y \right )$ means tensor contraction on all indexes in the
natural order. As $Y$ is in ${\cal D}_q^p$, its contraction with $T$ is a
locally integrable scalar function with compact support in the manifold and
therefore the above integral is well-defined. We will repeatedly use the
convention of distinguishing between a tensor field and the distribution it
defines by using an underline as before. As is seen in the last formula the
tensor distribution associated to a tensor field can act not only on
$C^{\infty}$ tensor fields with compact support but also on continuous ones. We
will consider the action of a tensor distribution always that this action can
be defined.

The components of a tensor distribution $\chi$ in a coordinate system are
scalar distributions $\chi^{\alpha_1\cdots\alpha_q}_{\beta_1\cdots\beta_p}$
defined by
\begin{eqnarray*}
\left < \chi^{\alpha_1\cdots\alpha_q}_{\beta_1\cdots\beta_p}, Y \right >
\equiv \left < \chi^q_p, Y dx^{\alpha_1}\otimes\cdots\otimes dx^{\alpha_p}
\otimes \frac{\partial}{\partial x^{\beta_1}}\otimes\cdots\otimes 
\frac{\partial}{\partial x^{\beta_q}} \right >
\end{eqnarray*}
where $Y$ is a function with compact support. With this definition it is not
difficult to prove the following expression
\begin{eqnarray*}
\left < \chi^q_p, Y^p_q \right > = \left < \chi^{\alpha_1\cdots\alpha_q}_{
\beta_1\cdots\beta_p},Y_{\alpha_1\cdots\alpha_q}^{\beta_1\cdots\beta_p}
\right > \enspace .
\end{eqnarray*}
The contraction of some indexes of a tensor distribution can be defined as the
tensor distribution whose components in a coordinate system are those obtained
by contracting the desired indexes of its components. This definition is seen
to be well defined and independent of the coordinate system.  Moreover, the
tensor product of a tensor distribution $\chi_p^q$ by a tensor field $T^r_s$
can be defined, (in general, only when this tensor field is $C^{\infty}$ but,
as we have already discussed, also in more general cases sometimes), as the
(p+s)-covariant (q+r)-contravariant tensor distribution acting as follows
\begin{eqnarray*}
\left < T^r_s\otimes\chi^q_p, Y^{s+p}_{r+q} \right > \equiv \left < 
\chi^q_p, \left ( T,Y\right )^p_q \right>
\end{eqnarray*}
where $(T,Y)^p_q$ is the element of ${\cal D}_q^p$ obtained by contracting $T$
with the first indexes of $Y$ in order.

We are now going to define covariant derivatives of tensor distributions. To
that aim, we have to consider Riemannian manifolds (or at least with a linear
connection satisfying Gauss theorem) such that the Christoffe, symbols are, in
principle, $C^{\infty}$ in each coordinate system. The definition of covariant
derivative of a tensor distribution generalizes the concept of covariant
derivative of tensor fields in the sense that for tensor distributions coming
from a tensor field it gives the tensor distributi$C^{\infty}$ tensor fields of
any order with compact support in the manifold. We denote by ${\cal D}^q_p$ the
subset of p-covariant q-contravariant tensor fields with compact support. Then,
p-covariant q-contravariant tensor distributions $\chi^q_p$ are defined as
linear and continuous functionals from ${\cal D}_q^p$ to the real numbers, that
is to say
\begin{eqnarray*}
\chi^q_p : {\cal D}^p_q & \rightarrow & {\rm I\!R} \\
Y^p_q & \rightarrow & \chi^q_p \left ( Y^p_q \right ) \equiv \left < \chi^q_p,
Y^p_q \right > \enspace .
\end{eqnarray*}
The sum of two tensor distributions of the same type and the product of a
tensor distribution by a real number can be defined in the usual way. With
these definitions the space of tensor distributions of a given type constitutes
a vector space. Given any locally integrable p-covariant q-contravariant tensor
field $T^q_p$ in an oriented manifold, we can define a p-covariant
q-contravariant tensor distribution associated to it as follows
\begin{eqnarray*}
{\underline{T}}^q_p : {\cal D}_q^p & \rightarrow & {\rm I\!R} \\ Y^p_q  &
\rightarrow & \left <{\underline{T}}^q_p, Y^p_q \right >\equiv
\int_{V_4} \left ( T,Y \right ) \mbox{\boldmath$\eta$}
\end{eqnarray*}
where $\left (T,Y \right )$ means tensor contraction on all indexes in the
natural order. As $Y$ is in ${\cal D}_q^p$, its contraction with $T$ is a
locally integrable scalar function with compact support in the manifold and
therefore the above integral is well-defined. We will repeatedly use the
convention of distinguishing between a tensor field and the distribution it
defines by using an underline as before. As is seen in the last formula the
tensor distribution associated to a tensor field can act not only on
$C^{\infty}$ tensor fields with compact support but also on continuous ones. We
will consider the action of a tensor distribution always that this action can
be defined.

The components of a tensor distribution $\chi$ in a coordinate system are
scalar distributions $\chi^{\alpha_1\cdots\alpha_q}_{\beta_1\cdots\beta_p}$
defined by
\begin{eqnarray*}
\left < \chi^{\alpha_1\cdots\alpha_q}_{\beta_1\cdots\beta_p}, Y \right >
\equiv \left < \chi^q_p, Y dx^{\alpha_1}\otimes\cdots\otimes dx^{\alpha_p}
\otimes \frac{\partial}{\partial x^{\beta_1}}\otimes\cdots\otimes 
\frac{\partial}{\partial x^{\beta_q}} \right >
\end{eqnarray*}
where $Y$ is a function with compact support. With this definition it is not
difficult to prove the following expression
\begin{eqnarray*}
\left < \chi^q_p, Y^p_q \right > = \left < \chi^{\alpha_1\cdots\alpha_q}_{
\beta_1\cdots\beta_p},Y_{\alpha_1\cdots\alpha_q}^{\beta_1\cdots\beta_p}
\right > \enspace .
\end{eqnarray*}
The contraction of some indexes of a tensor distribution can be defined as the
tensor distribution whose components in a coordinate system are those obtained
by contracting the desired indexes of its components. This definition is seen
to be well defined and independent of the coordinate system.  Moreover, the
tensor product of a tensor distribution $\chi_p^q$ by a tensor field $T^r_s$
can be defined, (in general, only when this tensor field is $C^{\infty}$ but,
as we have already discussed, also in more general cases sometimes), as the
(p+s)-covariant (q+r)-contravariant tensor distribution acting as follows
\begin{eqnarray*}
\left < T^r_s\otimes\chi^q_p, Y^{s+p}_{r+q} \right > \equiv \left < 
\chi^q_p, \left ( T,Y\right )^p_q \right>
\end{eqnarray*}
where $(T,Y)^p_q$ is the element of ${\cal D}_q^p$ obtained by contracting $T$
with the first indexes of $Y$ in order.

We are now going to define covariant derivatives of tensor distributions. To
that aim, we have to consider Riemannian manifolds (or at least with a linear
connection satisfying Gauss theorem) such that the Christoffel symbols are, in
principle, $C^{\infty}$ in each coordinate system. The definition of covariant
derivative of a tensor distribution generalizes the concept of covariant
derivative of tensor fields in the sense that for tensor distributions coming
from a tensor field it gives the tensor distribution associated to the usual
covariant derivative of the tensor field. This definition is
\begin{eqnarray*}
\left < \nabla \chi^q_p, Y^{p+1}_q \right > \equiv -\left < \chi ^q_p, (DY)^p_q
\right >
\end{eqnarray*}
where $\left (DY\right )^{\alpha_1\cdots\alpha_p}_{\beta_1\cdots\beta_q}=
\nabla_{\mu}Y^{\mu\alpha_1\cdots\alpha_p}_{\beta_1\cdots\beta_q}$.
With this definition, the components of the covariant derivative are the scalar 
distributions acting as
\begin{eqnarray}
\left < \nabla_{\mu} \chi^{\alpha_1\cdots\alpha_q}_{\beta_1\cdots\beta_p}, Y
\right > = -\left <\chi^{\alpha_1\cdots\alpha_q}_{\beta_1\cdots\beta_p},
\partial_{\mu}Y + \Gamma^{\rho}_{\rho\mu}Y\right>-\sum_{i=1}^{p}\left < 
\chi^{\alpha_1\cdots\alpha_q}_{\beta_1\cdots\beta_{i-1}\rho\beta_{i+1}\cdots
\beta_p}, \Gamma^{\rho}_{\beta_i\mu} Y\right > \nonumber\\
+ \sum_{j=1}^{q} \left<
\chi_{\beta_1\cdots\beta_p}^{\alpha_1\cdots\alpha_{j-1}\rho\alpha_{j+1}\cdots
\alpha_q}, \Gamma^{\alpha_j}_{\rho\mu}Y \right > .\hspace{35mm}\label{covdis}
\end{eqnarray}
In the case of a scalar distribution $\chi$ we have therefore
\begin{eqnarray*}
\left < \nabla_{\mu} \chi, Y \right > \equiv \left <\partial_{\mu} \chi, Y
\right > = - \left < \chi, \partial_{\mu}Y + \Gamma^{\rho}_{\rho\mu}Y \right >
\end{eqnarray*}
The last relations are written in the case of a Riemannian (or at least linear)
manifold but they also hold in the case of n-dimensional manifolds possessing a
$C^1$ volume form $\mbox{\boldmath$\eta$}$ by substituting 
$\Gamma^{\rho}_{\rho\mu}$ for $\Gamma_{\alpha}$ defined as
\begin{eqnarray*}
\partial_{\mu} \eta_{\beta_1\cdots\beta_n}\equiv \Gamma_{\mu}\eta_{\beta_1\cdots
\beta_n} \enspace .
\end{eqnarray*}
From now on, we will restrict ourselves to the case studied in chapter 3
whenever the preliminary junction conditions are satisfied, i.e. a spacetime
$V_4$ with a hypersurface $\Sigma$ such that the metric 
tensor is continuous but not differentiable across this hypersurface. Thus, the
Christoffel symbols do not exist at points of $\Sigma$ but they do everywhere
outside this hypersurface. First of all we will define the so-called
step Heaviside function of $\Sigma$ by
\begin{eqnarray}
\theta : V_4 & \rightarrow & {\rm I\!R}  \nonumber \\
\nonumber \\
&\theta \ = & \left \{ \begin{array}{c} \,1 \ {\rm in} \  V^{+} \\ a \ {\rm in}
\ \Sigma \\ \,0 \ {\rm in} \ V^{-} \\ \end{array} \right .  \label{teta}
\end{eqnarray}            
where $a$ is an arbitrary real number. This function is locally integrable and
therefore it defines a scalar distribution $\otheta$ in the natural way:
\begin{eqnarray*}
\left < \otheta, Y \right > = \int_{V^{+}} Y \mbox{\boldmath$\eta$}  
\end{eqnarray*}
The distribution $\otheta$ does not depend on the value of the function
$\theta$ on the hypersurface, and therefore we will not fix this value at the
moment.

In the case under consideration, the object $\Gamma_{\mu}$ given above does not
exist at the hypersurface but, in fact, this is not an obstacle to define the
partial derivative of some scalar distributions. In particular, let us consider
a scalar function $f$ which is discontinuous across $\Sigma$ but differentiable
in $V^{+}$ and $V^{-}$, and such that $f$ and its first derivatives have
definite limits on $\Sigma$ coming from both $V^{+}$ and $V^{-}$. If we call
$f^{+}$ its restriction to $V^{+}$ and analogously for $f^{-}$, it is trivial
to see that the distribution associated to that function, which exists because
$f$ is locally integrable, is
\begin{eqnarray*}
\underline{f}=f^{+}\cdot\otheta+f^{-}\cdot\left( \1-\otheta \right ) \enspace . 
\end{eqnarray*}
The partial derivative of this scalar distribution exists despite the
discontinuity of $\Gamma_{\mu}$ across $\Sigma$. Integrating by parts in
$V^{+}$ and $V^{-}$ we find
\begin{eqnarray}
\left <\nabla \underline{f},\vec{Y} \right>= 
\int_{V^{+}}Y^{\mu}\partial_{\mu}f^{+}\mbox{\boldmath$\eta$} + 
\int_{V^{-}}Y^{\mu}\partial_{\mu}f^{-} \mbox{\boldmath$\eta$} + 
\int_{\Sigma} \left [f \right ] Y^{\mu}d\sigma_{\mu} \label{partf}
\end{eqnarray}
where $d\sigma_{\mu}$ is oriented from $V^{-}$ towards $V^{+}$ and $\left[ f
\right ]$ is a scalar function defined on $\Sigma$, called step of $f$ at
$\Sigma$, and defined by
\begin{eqnarray*}
\forall q \in \Sigma \hspace{1cm} \ \left[ f \right ] (q) \equiv
 \mathop{\lim} \limits_{x \mathop \to \limits_{V^{+}}q} f^{+}(x) -
\mathop{\lim} \limits_{x \mathop \to \limits_{V^{-}}q} f^{-}(x) \enspace .
\end{eqnarray*}
We can rewrite all this by defining a one-covariant distribution
{\boldmath$\delta$} as
\begin{eqnarray*}
\left < \mbox{\boldmath $\delta$}, \vec{Y} \right >\equiv \int_{\Sigma}
Y^{\mu}d\sigma_{\mu}=\int_{\Sigma} Y^{\mu}n_{\mu} d\sigma \enspace .
\end{eqnarray*}
From equation (\ref{partf}), choosing $f^{+}=1$ and $f^{-}=0$ it is direct that
{\boldmath$\delta$} has an intrinsic definition as {\boldmath$\delta$}$=\nabla
\otheta$. This distribution can act on every vector field defined at least at
the points of the hypersurface which is locally integrable there. We can also
define a scalar distribution $\delta$ as follows
\begin{eqnarray*}
\left < \delta, Y \right >\equiv \int_{\Sigma} Y d\sigma \enspace .
\end{eqnarray*}
Of course, $\delta$ depends on the choice of the normal form ${\bf n}$ and we
have $\mbox{\boldmath$\delta$}={\bf n} \cdot \delta$ or, in components,
$\delta_{\mu}=n_{\mu}\cdot \delta$.  Therefore equation (\ref{partf}) can be
written as
\begin{eqnarray}
\partial_{\mu}\underline{f}= \partial_{\mu}f^{+}\, \cdot\otheta+
\partial_{\mu} f^{-}\, \cdot\left( \1-\otheta\right)+
\left[ f \right ] \cdot \delta_{\mu} \label{deriv} \enspace .
\end{eqnarray}

We are now going to define the connection and the Riemann tensor in the
manifold. $g$ being a continuous tensor across the hypersurface we can write it
as
\begin{eqnarray*}
g=\left( 1- \theta \right) g^{-} + \theta \, g^{+}
\end{eqnarray*}
independently of the value of $a$, and the tensor distribution associated to it
obviously is
\begin{eqnarray*}
\underline{g}=\left( \1- \otheta \right)\cdot g^{-} + \otheta\cdot g^{+} \enspace .
\end{eqnarray*}
As usual, we can define the connection symbols associated to the metric in the
manifold. We denote by ${\Gamma^{+}}^{\alpha}_{\beta\gamma}$ the Christoffel
symbols associated with $g^{+}$ and defined on $V^{+} \cup \Sigma$, by
${\Gamma^{-}}^{\alpha}_{\beta\gamma}$ those associated to $g^{-}$ and defined
on $V^{-} \cup \Sigma$, and finally by ${\underline{\Gamma}}^{\alpha}_{\beta
\gamma}$ the connection symbols associated with the whole metric distribution
$\underline{g}$ . The relation between them is obviously
\begin{eqnarray}
{\underline{\Gamma}}^{\alpha}_{\beta\gamma}=\left( \1- \otheta \right)\cdot
{\Gamma^{-}}^{\alpha}_{\beta\gamma}+
\otheta\cdot{\Gamma^{+}}^{\alpha}_{\beta\gamma} \enspace . \label{connexsym}
\end{eqnarray}
However, in order to be able to define covariant derivatives of tensor
distributions, and as is immediate from formula (\ref{covdis}), we need to know
the connection symbols not only as distributions but as functions in the
manifold as well. We choose these functions in a natural way from the
expression of the connection symbols as distributions. In consequence we have
\begin{eqnarray}
{\Gamma}^{\alpha}_{\beta\gamma}=\left( 1- \theta \right)
{\Gamma^{-}}^{\alpha}_{\beta\gamma}+\theta \,
{\Gamma^{+}}^{\alpha}_{\beta\gamma}
\enspace . \label{connexsymfunc}
\end{eqnarray}
Using the definition of the Riemann tensor
$R^{\alpha}_{\beta\lambda\mu}=\partial_{\lambda}\Gamma^{\alpha}_{\beta\mu}-
\partial_{\mu}\Gamma^{\alpha}_{\beta\lambda}+\Gamma^{\alpha}_{\lambda\rho}
\Gamma^{\rho}_{\beta\mu}-\Gamma^{\alpha}_{\mu\rho}\Gamma^{\rho}_{\beta\lambda}$
and treating this equation as a relation between distributions in the manifold,
we will find a relation between the Riemann tensor of $V_4$ and the Riemann
tensors defined from ${\Gamma^{+}}^{\alpha}_{\beta\gamma}$ and
${\Gamma^{-}}^{\alpha}_{\beta\gamma}$. In fact, from formulas (\ref{deriv}) and
(\ref{connexsym}) we have
\begin{eqnarray*}
\partial_{\mu}{\underline{\Gamma}}^{\alpha}_{\beta\lambda}=\left(\1-\otheta
\right)\cdot\partial_{\mu}{\Gamma^{-}}^{\alpha}_{\beta\lambda}+\otheta\cdot
\partial_{\mu}{\Gamma^{+}}^{\alpha}_{\beta\lambda}+\delta \cdot n_{\mu} \left[ 
\Gamma^{\alpha}_{\beta\gamma} \right ]
\end{eqnarray*}
and, given that ${\underline{\Gamma}}^{\alpha}_{\beta\gamma}$ are distributions
associated to functions so that the product
${\underline{\Gamma}}^{\alpha}_{\lambda\rho}
{\underline{\Gamma}}^{\rho}_{\beta\mu}$ is well defined
\begin{eqnarray*}
{\underline{\Gamma}}^{\alpha}_{\lambda\rho}{\underline{\Gamma}}^{\rho}_{\beta
\mu}=\left (\1-\otheta\right)
\cdot{\Gamma^{-}}^{\alpha}_{\lambda\rho}{\Gamma^{-}}^{\rho}_{\beta\mu}+
\otheta\cdot
{\Gamma^{+}}^{\alpha}_{\lambda\rho}{\Gamma^{+}}^{\rho}_{\beta\mu}
\end{eqnarray*}
where we have made use of $\otheta\cdot\otheta=\otheta$ and its consequences
$\otheta\cdot \left( \1- \otheta \right )=0$, $\left ( \1-\otheta
\right )\cdot\left (\1- \otheta \right ) = \left (\1 - \otheta \right )$,
we finally arrive at the following expression for the Riemann tensor
distribution
\begin{eqnarray}
{\underline{R}}^{\alpha}_{\beta\lambda\mu}=\left(\1-\otheta\right)
\cdot{R^{-}}^{\alpha}_{\beta\lambda\mu}
+\otheta\cdot {R^{+}}^{\alpha}_{\beta\lambda\mu}+\delta \cdot n_{\lambda}\left
[
\Gamma^{\alpha}_{\beta\mu} \right ] -\delta \cdot n_{\mu} \left [ 
\Gamma^{\alpha}_{\beta\lambda} \right ] .\label{dRiem}
\end{eqnarray}

\end{document}